\documentclass[preprint,11pt]{elsarticle}
\usepackage[T1]{fontenc}
\usepackage[english]{babel}
\usepackage{menukeys}
\usepackage{csquotes}
\usepackage{textcomp}
\usepackage{amsmath,amssymb}
\usepackage{mathtools}
\usepackage{tensor}
\usepackage{scalefnt}
\usepackage{graphicx}
\usepackage{listings}
\usepackage[tableposition=top]{caption}
\usepackage{booktabs}
\usepackage[final]{showkeys} 

\usepackage{acronym}
\usepackage{nth}
\usepackage{array}
\usepackage{enumitem}
\usepackage[hidelinks,bookmarksnumbered=true,plainpages=false,linktocpage=true]{hyperref}
\usepackage[capitalise]{cleveref}

\biboptions{sort&compress}

\newcommand{\fgfile}{\texttt{.fg}-file}
\newcommand{\editframe}{\code{EditFrame}}
\newcommand{\one}{one}
\newcommand{\two}{two}

\newcommand{\four}{four}
\newcommand{\feyngame}[1]{\textit{FeynGame#1}}
\newcommand{\qgraf}{\textit{qgraf}}
\newcommand{\order}[1]{{\mathcal{O}(#1)}}

\crefname{relation}{relation}{relations}
\creflabelformat{relation}{{\upshape(#2#1#3)}}
\definecolor{codegreen}{rgb}{.58,.69,.5}
\definecolor{codegray}{rgb}{.5,.5,.5}
\definecolor{codeblue}{rgb}{.35,.39,.6}
\definecolor{codered}{rgb}{.55,.3,.45}
\definecolor{backcolour}{rgb}{.95,.95,.95}
\newcommand\CODEstyle{\color{black}\ttfamily\scriptsize}
\lstdefinestyle{shell}{
  frame=single,
	language=bash,
    commentstyle=\color{codegreen},
    keywordstyle=\color{codeblue},
    numberstyle=\color{codegray}\ttfamily\tiny,
    basicstyle=\CODEstyle,
    breakatwhitespace=false,
    breaklines=true,
    postbreak=\mbox{\textcolor{codegray}{$\hookrightarrow$}\space},
    captionpos=b,
    keepspaces=true,
    numbers=none,
    numbersep=5pt,
    showspaces=false,
    showstringspaces=false,
    showtabs=false,
    tabsize=2,
    comment=[l]{\#},
}
\lstdefinestyle{qgraf}{
  backgroundcolor=\color{backcolour},
  language=Fortran,
  numberstyle=\color{codegray}\ttfamily\tiny,
    basicstyle=\CODEstyle,
    breakatwhitespace=false,
    breaklines=true,
    postbreak=\mbox{\textcolor{codegray}{$\hookrightarrow$}\space},
    captionpos=b,
    keepspaces=true,
    numbers=left,
    numbersep=5pt,
    showspaces=false,
    showstringspaces=false,
    showtabs=false,
    tabsize=2,
}

\newcounter{bla}

\journal{Computer Physics Communications}

\newpageafter{abstract}

\newcommand*{\abbrev}[1]{{\scalefont{.9}#1}}
\newcommand*{\citere}[1]{Ref.~\cite{#1}}
\newcommand*{\citeres}[1]{Refs.~\cite{#1}}
\newcommand*{\code}[1]{\texttt{#1}}

\newcounter{notecount}

\newcommand{\myacrodef}[3]{\acrodef{#2}{#3}\newcommand{#1}{\ac{#2}}}
\myacrodef{\gui}{GUI}{graphical user interface}
\myacrodef{\pdf}{PDF}{leading-order}
\acused{PDF}
\myacrodef{\lo}{LO}{leading-order}
\myacrodef{\nlo}{NLO}{next-to-\lo}
\myacrodef{\nnlo}{NNLO}{next-to-\nlo}
\myacrodef{\onepi}{1PI}{\one-particle-irreducible}
\myacrodef{\sm}{SM}{Standard Model}
\myacrodef{\qft}{QFT}{Quantum Field Theory}
\acrodefplural{QFT}{Quantum Field Theories}

\myacrodef{\QED}{QED}{Quantum Electrodynamics}
\myacrodef{\qcd}{QCD}{Quantum Chromodynamics}
\myacrodef{\eft}{EFT}{Effective Field Theory}
\acrodefplural{EFT}{Effective Field Theories}

\myacrodef{\ibp}{IbP}{integration-by-parts}

\usepackage{fancyhdr}
\newcommand{\RHheaderline}{\textsf{TTK-24-56~/~P3H-24-096---~January~2025}}
\fancypagestyle{pprintTitle}
{
  
  \fancyhead[R]{\RHheaderline}
}
\begin{document}

\begin{frontmatter}

  \title{\includegraphics[width=.15\textwidth]{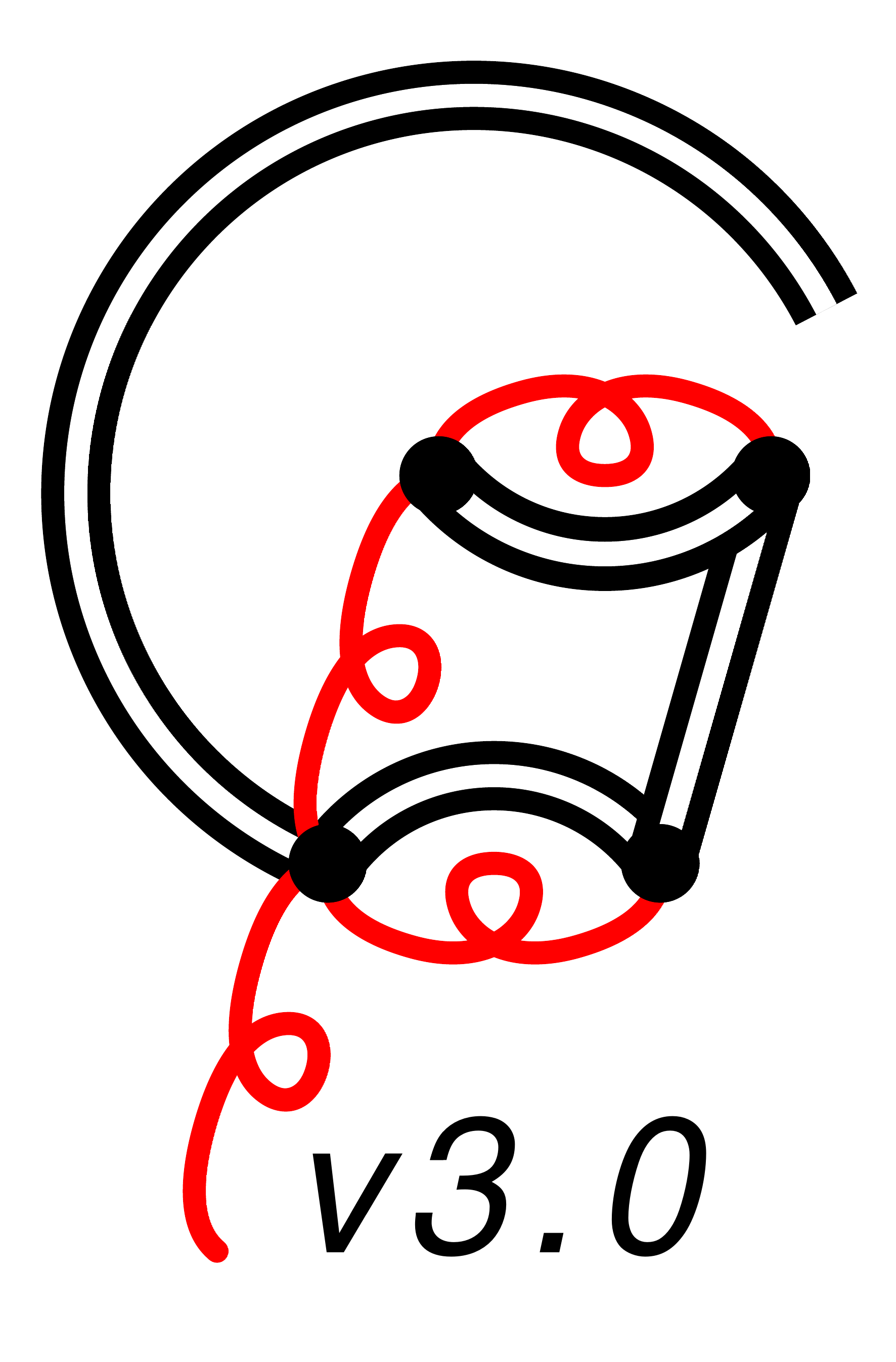}\\[1em]
  FeynGame 3.0}

\author{Lars~B\"undgen}
\author{Robert~V.~Harlander}
\author{Sven~Yannick~Klein}
\author{Magnus~C.~Schaaf}

\address{Institute for Theoretical Particle Physics and Cosmology,\\ RWTH Aachen University, 52056 Aachen, Germany}

\begin{abstract}
  A major update of the program \feyngame{} is introduced.  One of its main
  new functionalities is to visualize Feynman graphs generated by \qgraf. The
  \qgraf\ output can be either pasted into the \feyngame\ canvas for
  individual graphs, or the whole \qgraf\ output file can be processed. In
  addition, a number of new features and improvements have been implemented
  into \feyngame{-3.0} in order to further facilitate the efficient drawing of
  Feynman diagrams in publication quality.  \feyngame\ is freely
  available
  \begin{itemize}
  \item 
    as \texttt{jar} or MacOS \texttt{app} file from\\
    \url{https://web.physik.rwth-aachen.de/user/harlander/software/feyngame}
  \item as source code from \url{https://gitlab.com/feyngame/FeynGame}
  \end{itemize}
\end{abstract}

\begin{keyword}
Feynman diagrams, graphical user interface, spring layout algorithm
\end{keyword}

\end{frontmatter}

\noindent \textbf{PROGRAM SUMMARY}

\begin{small}
\noindent
{\em Program Title\/:}  
\feyngame{-3.0}

{\em Developer's repository link\/:}
\url{https://gitlab.com/feyngame/FeynGame}

{\em Licensing provisions\/:}
GNU General Public License 3 (GPLv3)

{\em Programming language\/:}                    
\texttt{Java}

{\em Nature of problem:} Efficient drawing of Feynman diagrams for
presentations and publications; automatic visualization of computer-generated
Feynman graphs~\cite{qgraf}; playful elementary introduction to the
concept of Feynman diagrams.

{\em Solution method:}
Graphical interface based on version~1 of
\feyngame~\cite{feyngame}; spring-layout algorithm~\cite{spring} for the
automatic positioning of vertices.

\end{small}

\section{Introduction}

Modern perturbative calculations often involve thousands, if not millions of
Feynman graphs. Obviously, this requires a large degree of automation, with
human intervention kept at very few and highly localized interaction
points. In particular, Feynman graph generation, insertion of Feynman rules
and many of the subsequent algebraic operations are typically part of a fully
automated workflow.

Visual representations of the graphs, i.e.\ the associated diagrams, seem to
be irrelevant in a modern calculation. While the role of Feynman diagrams as
communicative tool remains undebated, be it at the scientific or educative
level, the typical toolchain of the practitioner does not require any drawing
of the diagrams.

However, this is only true if the toolchain works flawlessly. But at the
frontier of scientific research, this is usually not the case. Quite
generally, the number of relevant physical problems that can be solved
completely with existing tools is finite, and thus they will be solved by
somebody at some point. But progress in science happens by solving as-of-yet
unsolvable problems. It requires to go beyond the limits of existing tools and
methods. In this process, it frequently happens that a toolchain fails at a
certain point, and the process of ``debugging'' begins.

Quite a common situation in a perturbative calculation is that, after
calculating $N\sim \order{10^4}$ Feynman graphs, the workflow unexpectedly
terminates during the calculation of graph number $N+1$, for example due to
some memory overflow. This immediately poses the question about what
distinguishes this graph from the previous $N$ graphs. It is here where one
may want to actually \textit{see} the corresponding diagram.

One of the most popular Feynman graph generators is
\qgraf~\cite{Nogueira:1991ex,Nogueira:2021wfp}, mostly due to its
computational speed. While its output is human readable and highly
customizable \code{ASCII} code, it does not provide the corresponding visual
diagrams. Most researchers working in the field will have found themselves
manually connecting dots and lines on a piece of paper, aiming to translate
the \qgraf\ output to a diagram. For complicated diagrams, this usually is a
two-step process, where the first step involves figuring out the topology of
the diagram, and in the second step, one tries to disentangle crossed lines by
moving around vertices and curving lines.

In fact, already the original version of \feyngame{} was very helpful in this
respect because it literally allows one to move vertices around in a
topology-preserving manner~\cite{Harlander:2020cyh}. Originally developed as
an educational tool for high-school and undergraduate students, \feyngame{}
has turned out to be quite useful also for preparing high-quality Feynman
diagrams for scientific publications and presentations in a very efficient
way.

While the original release already had most of the distinctive features, the
recently published version~2.1 has been improved in several respects,
including full \LaTeX\ support for line and vertex labels. One of its most
significant new features was to produce the mathematical form of the amplitude
from the diagram drawn on the canvas. 

With the current release of \feyngame{}, we include the functionality of
converting \qgraf\ output to visual diagrams. It relies on the spring layout
algorithm to draw the diagrams and allows for various adjustments of the
associated parameters.  Aside from this, \feyngame{-3.0} includes a large
number of new features and improvements which we will describe in this
paper. Among them is the preview of diagrams in the file chooser, a
double-grid, and the possibility to convert diagrams to different models.

The remainder of the paper is structured as follows. After briefly reviewing
the underlying philosophy of \feyngame, we will explain in detail the
\qgraf\ visualization functionality in \cref{sec:qgraf}. Even though it was
already included in \feyngame{-2.1} and briefly described in
\citere{Harlander:2024qbn}, we will describe the feature of producing the
algebraic form of the amplitude associated with a particular diagram in
\cref{sec:amplitude}. Other improvements and new features will be briefly
introduced in \cref{sec:improvements}. While the \gui\ of \feyngame\ is
one of its central elements that allow for an efficient drawing of Feynman
diagrams, it is often useful to be able to perform certain operations on
existing diagrams (export, conversion, etc.) without opening the \gui. This is
why \feyngame\ is also equipped with a command-line mode, whose most important
commands are described in \cref{sec:commandline}. Conclusions and an outlook
are presented in \cref{sec:conclusions}.

\section{The \feyngame{} philosophy}\label{sec:philo}

\begin{figure}
  \begin{center}
    \begin{tabular}{c}
    \begin{tabular}{cc}
      \raisebox{0em}{%
          \includegraphics[%
            clip,width=.37\textwidth]%
                          {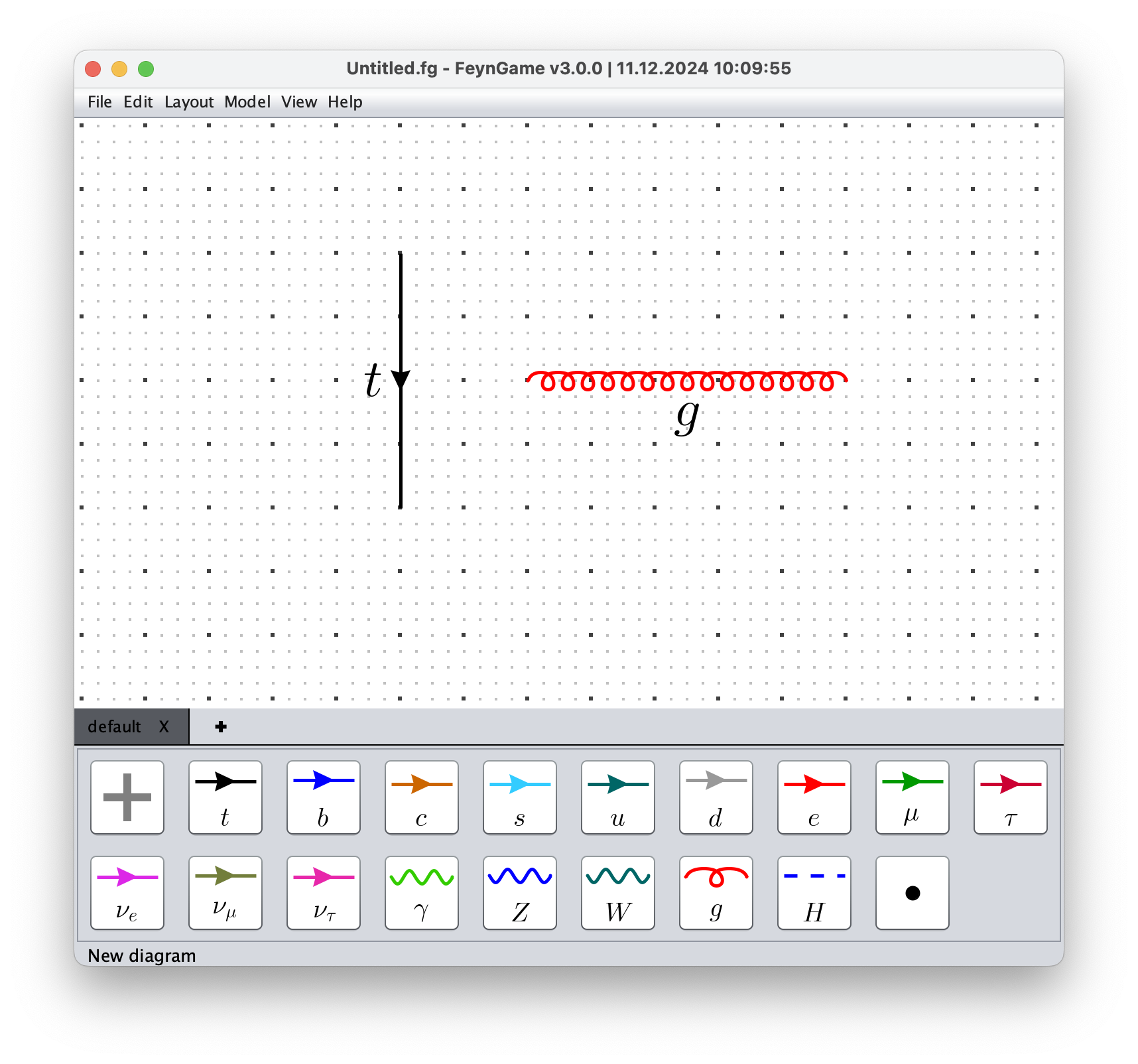}}
      &
      \raisebox{0em}{%
          \includegraphics[%
            clip,width=.37\textwidth]%
                          {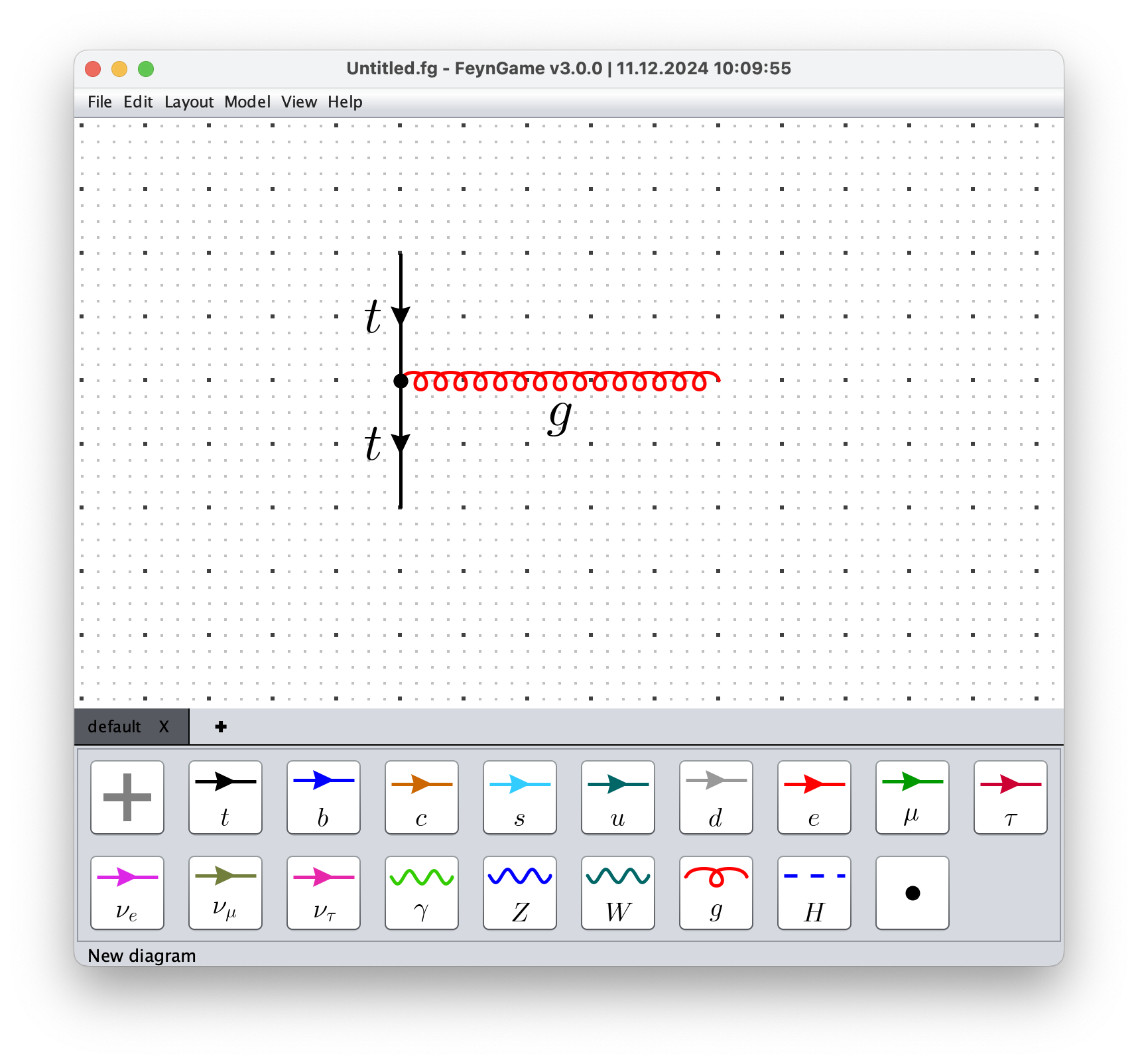}}\\[-1em]
      (a) & (b)\\[1em]
      \raisebox{0em}{%
          \includegraphics[%
            clip,width=.37\textwidth]%
                          {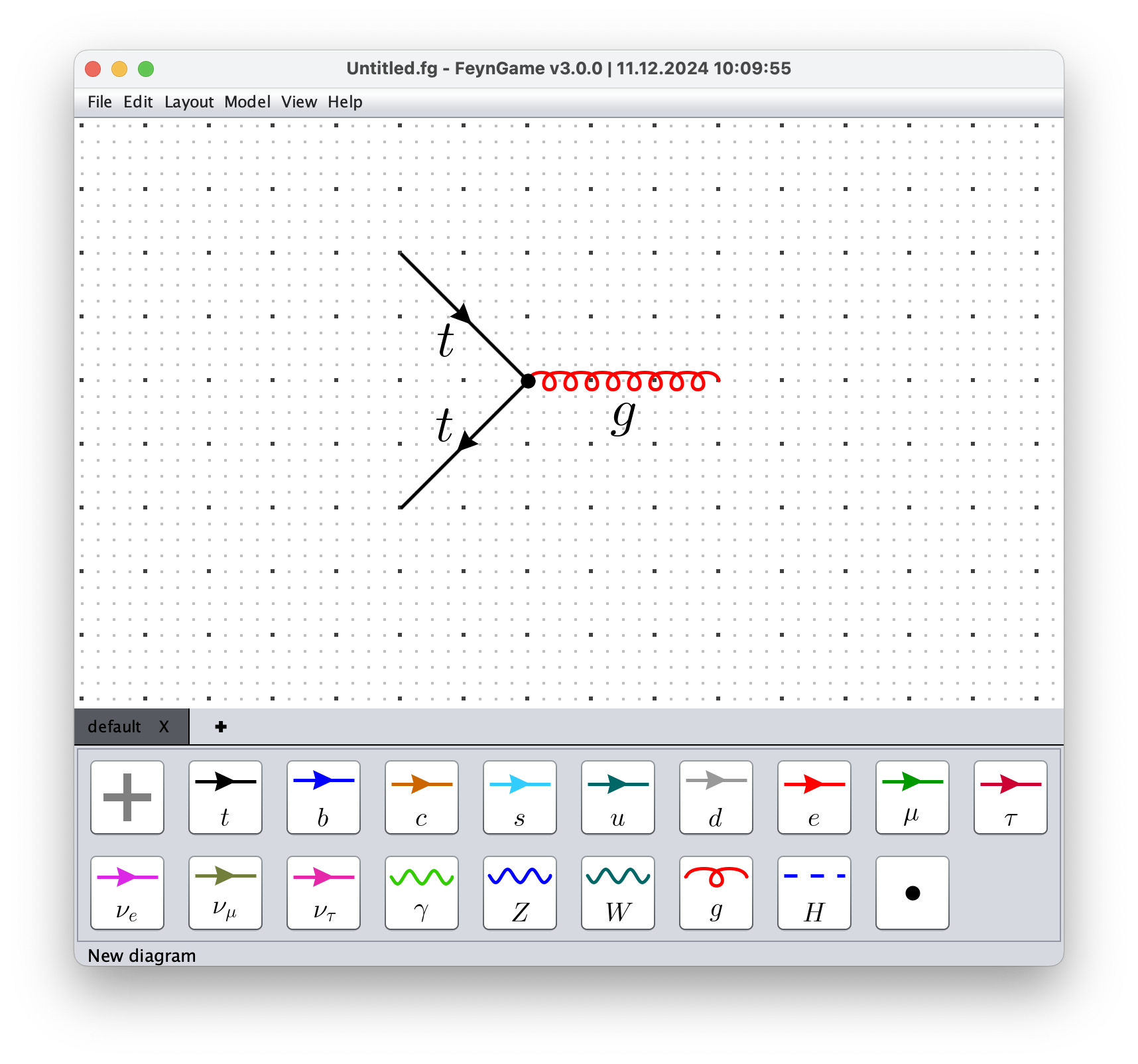}} &
      \raisebox{0em}{%
          \includegraphics[%
            clip,width=.37\textwidth]%
                          {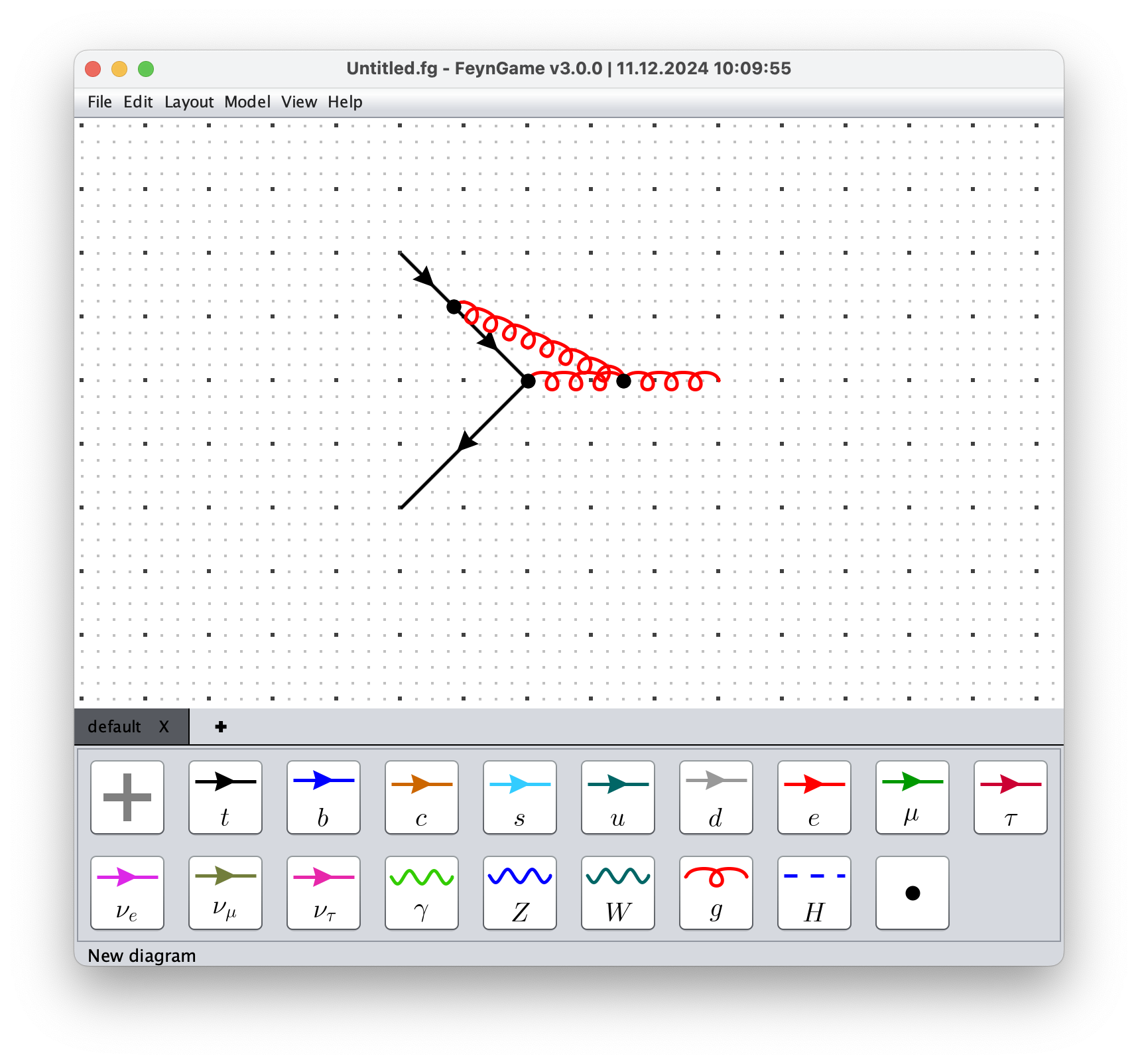}}\\[-1em]
    (c) & (d)
    \end{tabular}\\[1em]
      \raisebox{0em}{%
          \includegraphics[%
            clip,width=.37\textwidth]%
                          {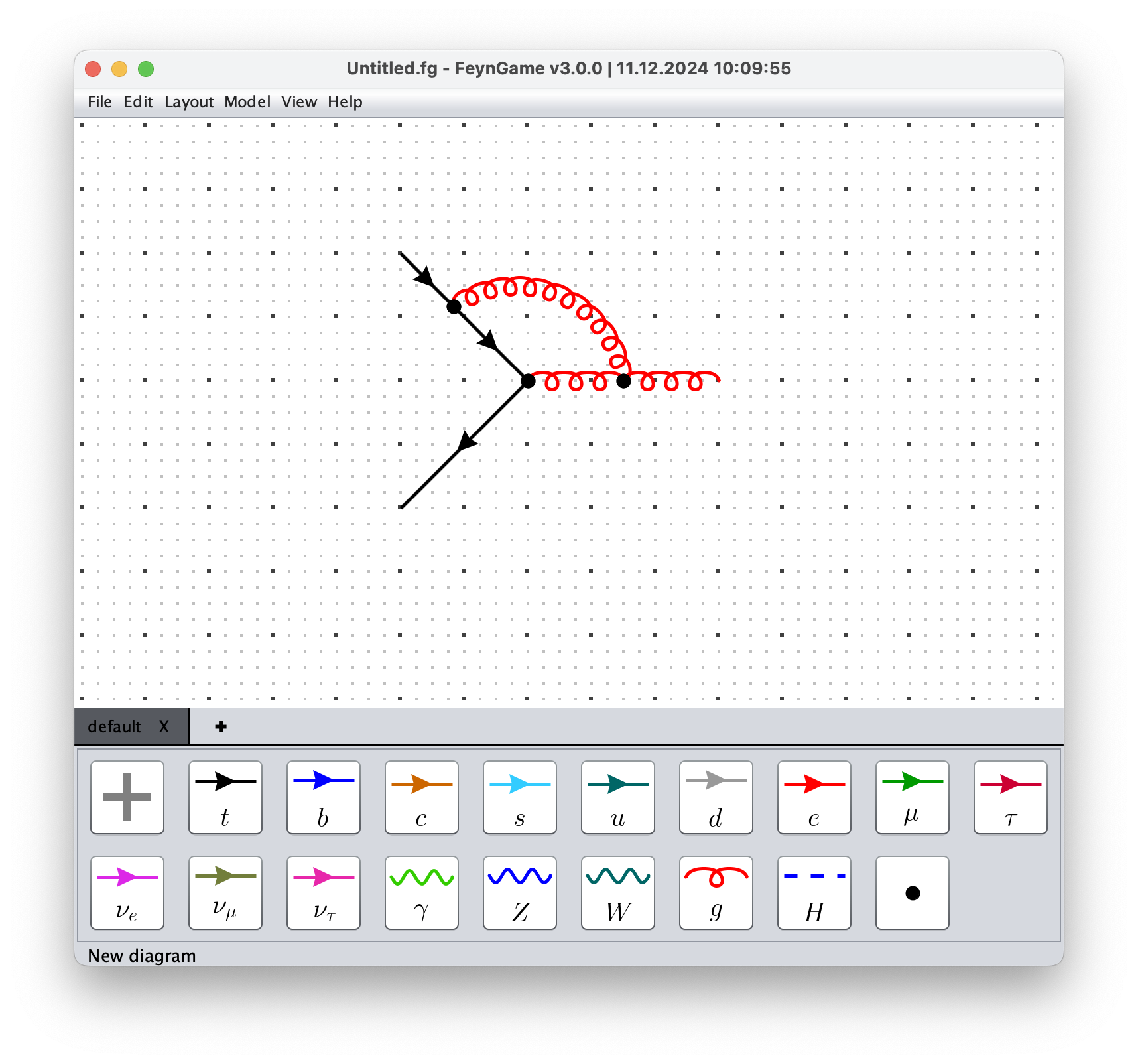}}\\[-1em]
      (e)\\[1em]
    \end{tabular}
    \parbox{.9\textwidth}{
      \caption[]{\label{fig:philosophy}\sloppy Screenshots of
        \feyngame\ illustrating its basic philosophy. They show the main
        window of \feyngame\ with the \textit{canvas} (upper part) and the
        \textit{current model} (lower part), represented by tiles containing
        the particles of the Standard Model. (a)~A top-quark $t$ and a gluon
        line $g$ are drawn on the canvas; (b)~a $t\bar tg$ vertex was created
        by moving the gluon line close to the top-quark line; (c)~the shape of
        the diagram was changed by clicking-and-dragging the vertex point to
        the right; (d)~an additional gluon line was attached to form a loop
        diagram; (e)~the gluon line was curved using the track pad (mouse
        wheel).}}
  \end{center}
\end{figure}

\feyngame's philosophy differs in various aspects from other programs designed
for drawing Feynman diagrams. Its most important characteristics are:
\begin{itemize}
\item \feyngame{} is model-based, meaning that it ``knows'' the
  particles\footnote{We use ``particle'' and ``field'' interchangably.}  and
  their interaction vertices of the underlying model (the \sm\ by
  default). Each particle can be assigned its unique line style
  (plain/wave/spiral/..., dashing, thickness, color, directed/undirected,
  label, etc.). Instead of drawing a generic line and modifying its
  attributes, one simply draws the line of a specific particle, for example a
  gluon or a top quark, and it will already have the correct attributes; see
  \cref{fig:philosophy}\,(a) 
\item \feyngame{} automatically creates vertices if the end of line~A is
  dragged to the vicinity of line~B anywhere along line~B. The two parts of
  line~B separated by the vertex are then treated by \feyngame{} as two distinct
  lines; see \cref{fig:philosophy}\,(b) and (d).
\item When moving a vertex on the canvas, lines attached to it will be dragged
  along. This makes it very easy to adjust the shape of a Feynman diagram; see
  \cref{fig:philosophy}\,(c).
\item Lines can be moved/rotated/stretched/shrunk using the mouse in a very
  intuitive way.
\item For \feyngame{}, straight or curved lines are represented by the same
  type of object.  By default, the user draws a straight line and subsequently
  applies a curvature with the mouse wheel (or some other suitable input
  device) if desired, see \cref{fig:philosophy}\,(e). A line can even be
  curved into a full circle.
\item Many operations are accessible in various ways, for example via the
  menu, the keyboard/mouse, or the so-called \editframe.
\end{itemize}
We recommend that the reader who is not yet familiar with \feyngame{} to
download it from the
\href{https://web.physik.rwth-aachen.de/user/harlander/software/feyngame/}{\feyngame{}
  website} and start drawing. Its general functionality should be sufficiently
intuitive to draw standard Feynman diagrams without reading any
instructions. In order to explore all features of \feyngame, we recommend to
consult \citeres{Harlander:2020cyh,Harlander:2024qbn}, and of course the
remainder of the current paper.

\section{Importing \qgraf\ output}\label{sec:qgraf}

One of the most significant new features of \feyngame{-3.0} is its capability
to visualize the output of \qgraf\ in diagrammatical form. \qgraf\ is a
software tool to generate Feynman graphs in a very efficient
way~\cite{Nogueira:1991ex,Nogueira:2021wfp}. For example, it takes of the
order of a second to generate all 10,584 \onepi\ graphs for the process
$e^+e^-\to \gamma\gamma$ at \four-loop level within \QED. However, it does not
provide a visual representation of the generated diagrams. This section
describes how \feyngame{-3.0} supplies this capability.

\subsection{Basic functionality}

\begin{figure}
  \begin{center}
    \begin{tabular}{c}
      \raisebox{0em}{%
          \includegraphics[%
            width=.3\textwidth]%
                          {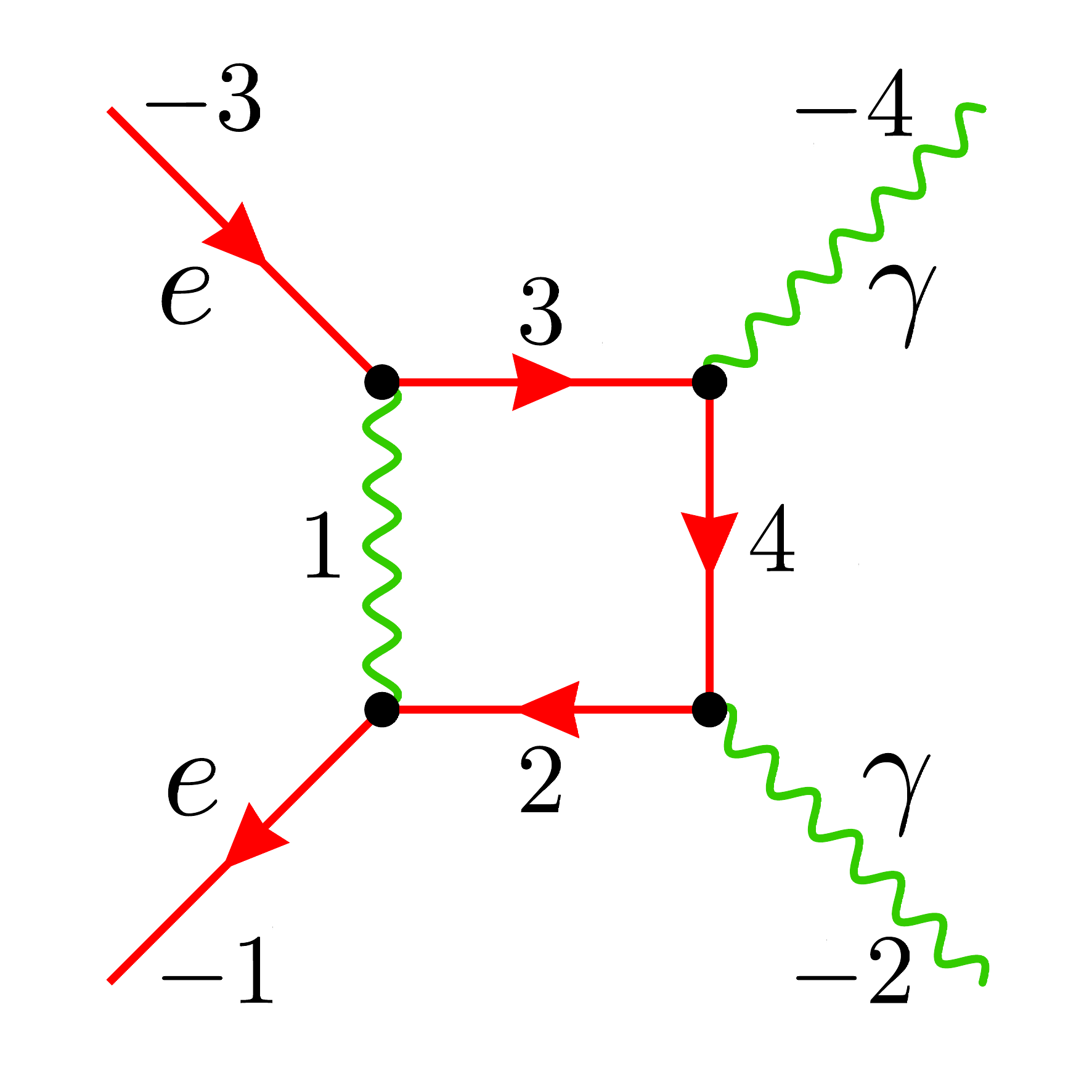}}
    \end{tabular}
    \parbox{.9\textwidth}{
      \caption[]{\label{fig:eegamgam-1l}\sloppy The diagram corresponding to
        the graph of \cref{lst:eegamgam-1l}. The numbers on the lines
        correspond to the line labels of \qgraf\ in \cref{lst:eegamgam-1l}.}
    }
  \end{center}
\end{figure}

Given a specific interaction model like \qcd\ or the \sm, \qgraf\ writes the
set of diagrams that contribute to a specific process to an \texttt{ASCII}
output file. The format of the output is highly adjustable by the user via a
so-called style file. The \qgraf\ distribution provides a number of sample
style files. For example, using \code{form.sty}, the \qgraf\ output for a
typical \one-loop graph for the process $e^+e^-\to \gamma\gamma$ reads
\begin{lstlisting}[%
    style=qgraf,
    caption={\qgraf\ output for a \one-loop Feynman
      graph contributing to the process $e^+e^-\to \gamma\gamma$.},
    label=lst:eegamgam-1l
  ]
*--#[ d1:
*
     1
    *vx(E(2),e(-1),ph(1))
    *vx(E(-3),e(3),ph(1))
    *vx(E(4),e(2),ph(-2))
    *vx(E(3),e(4),ph(-4))
*
*--#] d1:
\end{lstlisting}
This indicates that the diagram contains four \QED\ vertices \texttt{vx(...)}.
The symbols \code{e}, \code{E}, and \code{ph} denote the particles $e^-$,
$e^+$, and $\gamma$, respectively, and the integers in brackets label the
different lines of the diagram (positive: propagators, negative: external
particles). The reader may easily verify that the corresponding diagram is
given by the one of \cref{fig:eegamgam-1l}.

\begin{figure}
  \begin{center}
    \begin{tabular}{c}
      \raisebox{0em}{%
        \includegraphics[%
          angle=0,
            width=.9\textwidth]%
                          {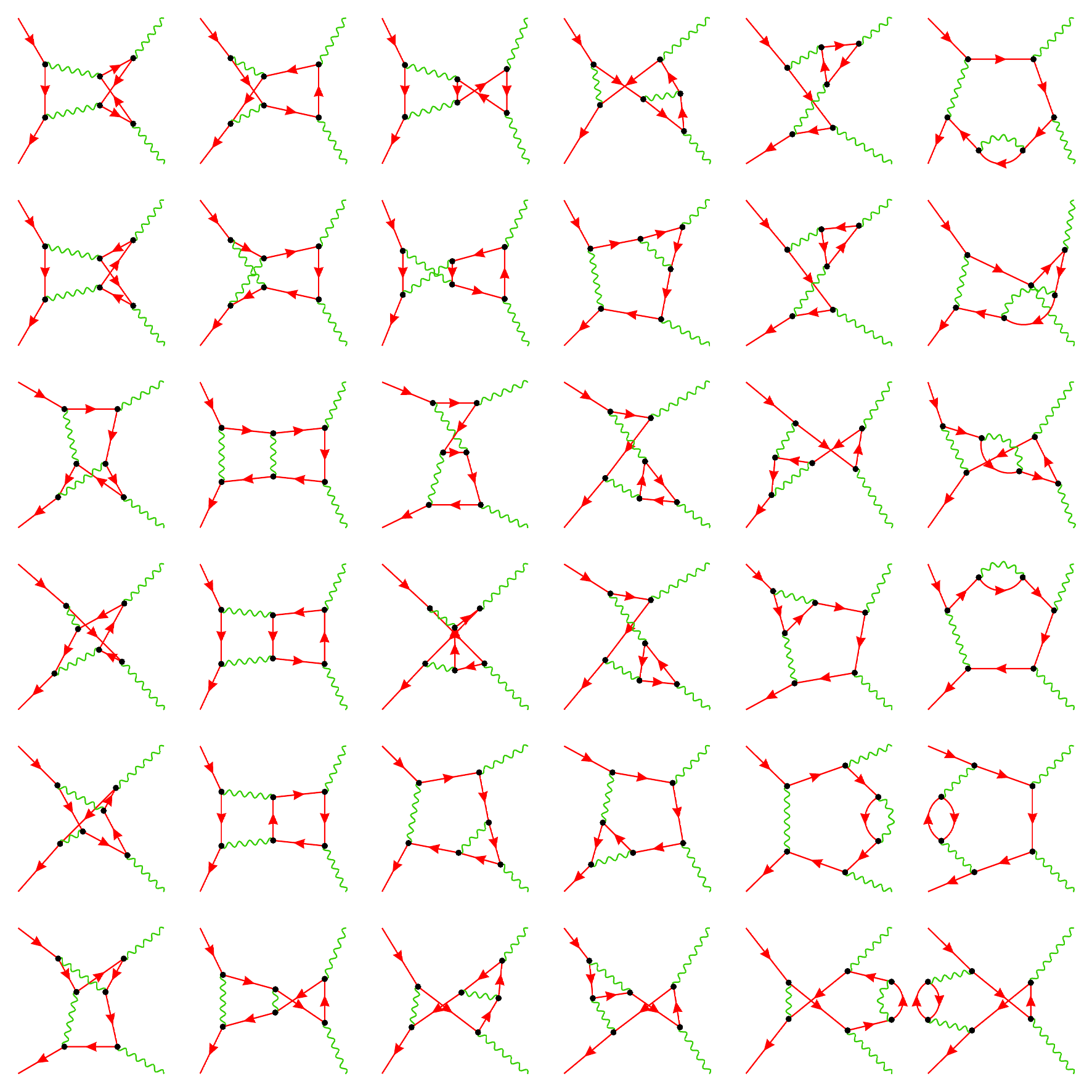}}
    \end{tabular}
    \parbox{.9\textwidth}{
      \caption[]{\label{fig:eegamgam-2l-all}\sloppy All \two-loop
        \onepi\ Feynman diagrams contributing to the process $e^+e^-\to
        \gamma\gamma$ in \QED\ as drawn by \feyngame{} from the corresponding
        \qgraf\ output.  }}
  \end{center}
\end{figure}

In fact, this image was obtained by a simple copy-paste operation of
\cref{lst:eegamgam-1l} into the \feyngame{} canvas, without any further manual
adjustments. To be able to better appreciate the power of this feature, let us
consider a \two-loop diagram of the same process:
\begin{lstlisting}[%
    style=qgraf,
    caption={\qgraf\ output for a \two-loop Feynman
      graph contributing to the process $e^+e^-\to \gamma\gamma$.},
    label=lst:eegamgam-2l
  ]
*--#[ d...:
*
     -1
    *vx(E(1),e(-1),ph(2))
    *vx(E(-3),e(1),ph(3))
    *vx(E(5),e(4),ph(-2))
    *vx(E(4),e(6),ph(-4))
    *vx(E(7),e(5),ph(2))
    *vx(E(6),e(7),ph(3))
*
*--#] d...:
\end{lstlisting}
We challenge the reader to find the corresponding diagram in
\cref{fig:eegamgam-2l-all}, which shows all \two-loop \onepi\ diagrams
contributing to $e^+e^-\to \gamma\gamma$ in \QED. The figure was produced by
\feyngame{} from the corresponding \qgraf\ output, again without any further
manual manipulation. In this case, the whole \qgraf\ output file was imported
into \feyngame, and the resulting diagrams were exported to \abbrev{PDF}. This
can be done using \menu[,]{File,Import qgraf}, and subsequently
\menu[,]{qgraf,Export all diagrams}.

\subsection{\qgraf\ style}\label{sec:qgraf-style}

The specific form of the \qgraf\ output is highly adjustable by what we call
the ``\qgraf\ style file''.  In particular, the user may choose to encode a
Feynman graph in terms of its vertices, as it is done in the examples of
\cref{lst:eegamgam-1l,lst:eegamgam-2l}, or in terms of its propagators (or
both). For example, using the style file
\code{qgraf-tapir.sty}~\cite{Gerlach:2022qnc}, the diagram in
\cref{lst:eegamgam-1l} would be output by \qgraf\ as
\begin{lstlisting}[style=qgraf, caption={The graph represented by
    \cref{lst:eegamgam-1l,fig:eegamgam-1l} using the \code{qgraf-tapir.sty}
    style file.}, label=lst:eegamgam-1l-tapir,
  firstline=15,lastline=37]
{
diagram            1
pre_factor         (+1)*1

number_propagators 4
number_loops       1
number_legs_in     2
number_legs_out    2


external_leg       q1|1|e
external_leg       q2|2|E


external_leg       q3|3|ph
external_leg       q4|4|ph


momentum           p1|2,1|ph,ph
momentum           p2|1,3|E,e
momentum           p3|4,2|E,e
momentum           p4|3,4|E,e
}
\end{lstlisting}

For this reason, \feyngame{-3.0} needs to know the style file which was used
to produce the output.  It comes with a number of pre-defined styles, among
them \texttt{form.sty} and \texttt{qgraf-tapir.sty}. User-specific style files
can be added via \menu[,]{File,Settings,Qgraf Options}. \feyngame{} will
remember the style file also for future sessions, until it is removed by the
user via the same interface.

If a single graph from the \qgraf\ output file is pasted into the canvas,
\feyngame{} will try to draw it on the basis of any of the known style
files. On the other hand, if the user imports the \qgraf\ output via
\menu[,]{File,Import qgraf}, one needs to select the \qgraf\ style file from
\feyngame's list of known styles, or add a new one.

\subsection{Synchronizing the \qgraf\ and the \feyngame{} model
  file}\label{sec:sync}

In order for \feyngame{} to identify the particles in the \qgraf\ output file,
they have to be part of the \feyngame{} model file. This section will describe
how to achieve this.

Aside from the style file (see \cref{sec:qgraf-style}), \qgraf\ requires two
input files. One of them, typically referred to as \code{qgraf.dat}, specifies
the type of diagrams to be generated: initial and final state, number of
loops, topological restrictions, etc. What concerns us here, however, is the
so-called \qgraf\ model file. It defines the underlying model (or theory),
such as \QED, \qcd, or the \sm. For example, \QED\ with a single electron
could be defined as 
\begin{lstlisting}[style=qgraf, caption={\qgraf\ model file
    for \QED.}, label=lst:qgraf_qed]
   [ph, ph, + ]
   [e, E, - ]
   [E, e, ph ]
   
\end{lstlisting} Lines~1~and~2 specify the
particle content, including the spin-statistics (\code{+} for bosons, \code{-}
for fermions), and line~3 defines the only vertex of theory. For more details
and options concerning the \qgraf\ model file, we refer the reader to
\citeres{Nogueira:1991ex,Nogueira:2006pq}. The crucial observation in the
current context is that each particle is associated with a unique,
user-defined identifier. In \cref{lst:qgraf_qed}, this is \code{ph} for the
photon, \code{e} for the electron, and \code{E} for its adjoint.  These
identifiers will also be used by \qgraf\ to encode the generated Feynman
diagrams, see \cref{lst:eegamgam-1l,lst:eegamgam-2l,lst:eegamgam-1l-tapir}.

The basic structure of the \feyngame{} model file is very close to the one of
\qgraf. For example, the default \feyngame{} model file contains the lines
(among many others)
\begin{lstlisting}[style=qgraf]
  [ph, ph, WAVE, color=ff00ff00, stroke=4.0, dash=false, <...> ]
  [e, E, PLAIN, color=ffff0000, stroke=4.0, dash=false, <...> ]
\end{lstlisting}
where \code{<...>} denotes more line specifications. Also here, the first two
elements in each square bracket are identifiers for the particles. The fact
that they coincide with the identifiers used by \qgraf\ allows \feyngame{} to
associate the respective line style to the diagrams in the \qgraf\ output.

This means that, in order to achieve the most satisfactory results when
displaying a \qgraf\ output file in \feyngame, the user is asked to
synchronize their \feyngame{} and \qgraf\ model file. Since the latter is
typically part of an existing toolchain, the user may be reluctant to modifiy
the identifiers here. We therefore recommend to adjust the identifiers in the
\feyngame{} model file instead. This can be done in several ways:
\begin{itemize}
\item If the user's \qgraf\ model file represents the \sm\ or a subset
  thereof, one may directly edit the default \feyngame{} model file in one's
  favorite editor and simply adjust the particle identifiers. The location of
  the default model file is platform dependent:
  \begin{itemize}
  \item MacOS: \verb#$HOME/Library/Preferences/FeynGame/default.model#
  \item Linux: \verb#$HOME/.config/FeynGame/default.model#
  \item Windows: \verb#%APPDATA%\FeynGame\default.model#
  \end{itemize}
  The location is also reported at the bottom line of the \feyngame{} main
  window when loading the default model file with \menu[,]{Model,Load default
    model} (see also \cref{sec:improvements},
  \textit{\hyperlink{pos:action}{Action messages}}).
\item Otherwise, the user may convert their \qgraf\ model file into a
  \feyngame{} model file using \menu[,]{Model,Import qgraf model}. This
  automatically generates a unique line style for each particle which the user
  can then edit from within \feyngame{} with the help of the \editframe\
  (see \citere{Harlander:2020cyh,Harlander:2024qbn} for details), and
  subsequently save to disk, or even set as the default model file. Of course,
  the generated \feyngame{} model file can also be edited in one's favorite text
  editor.
\item If the Feynman graph imported from \qgraf\ involves a particle whose
  identifier does not appear in the \feyngame{} model file, \feyngame{} will
  automatically generate a unique line style for that particle. It will prompt
  the user to decide whether the new particle should be added to the current
  model. In this way, one can successively create a \feyngame{} model file
  which is suitable for the process under consideration.
\end{itemize}

\subsection{The drawing algorithm}\label{sec:drawing}

As already pointed out above, \qgraf{} only provides extended information
about the topological structure of the Feynman graphs. In mathematical terms,
the essential information can be described as a \textit{directed
  pseudograph}. In order to display this graph as a diagram and place the
vertices and lines in an optimal way, \feyngame{} adopts the so-called spring
layout algorithm, which is also used by the \LaTeX\ package
\textit{TikZ-Feynman}\,\cite{Ellis:2016jkw}.  It belongs to the class of
force-directed graph drawing algorithms which were first described in
\citere{eades84}.

The idea behind this type of algorithm is to describe the graph as a physical
system whose minimal energy state corresponds to the desired graph layout. A
vertex is treated as a charged mass point and a propagator is treated as a
spring. The charges of all vertices are equal, so there is a repulsive force
between any two vertices.  The scale of the spacing between the vertices is
controlled by a length parameter $\ell$: The springs that model the
propagators are tuned such that, when combined with the electrical force, they
produce a net attractive force between vertices for distances $d>\ell$ and a
net repulsive force for distances $d<\ell$. This keeps vertices which are
directly connected by a spring close to one another. The exact configuration
of spring and electrical forces is described in
\citere{Fruchterman1991GraphDB}.  Starting from a random distribution of the
vertices, the system is then evolved towards its minimum energy state in
discrete timesteps. The simulation is stopped after a certain number of
iterations, and the resulting physical system is used to set the vertex
positions in the Feynman diagram.

\begin{figure}
  \begin{center}
    \begin{tabular}{cc}
          \includegraphics[%
            width=.26\textwidth]%
                          {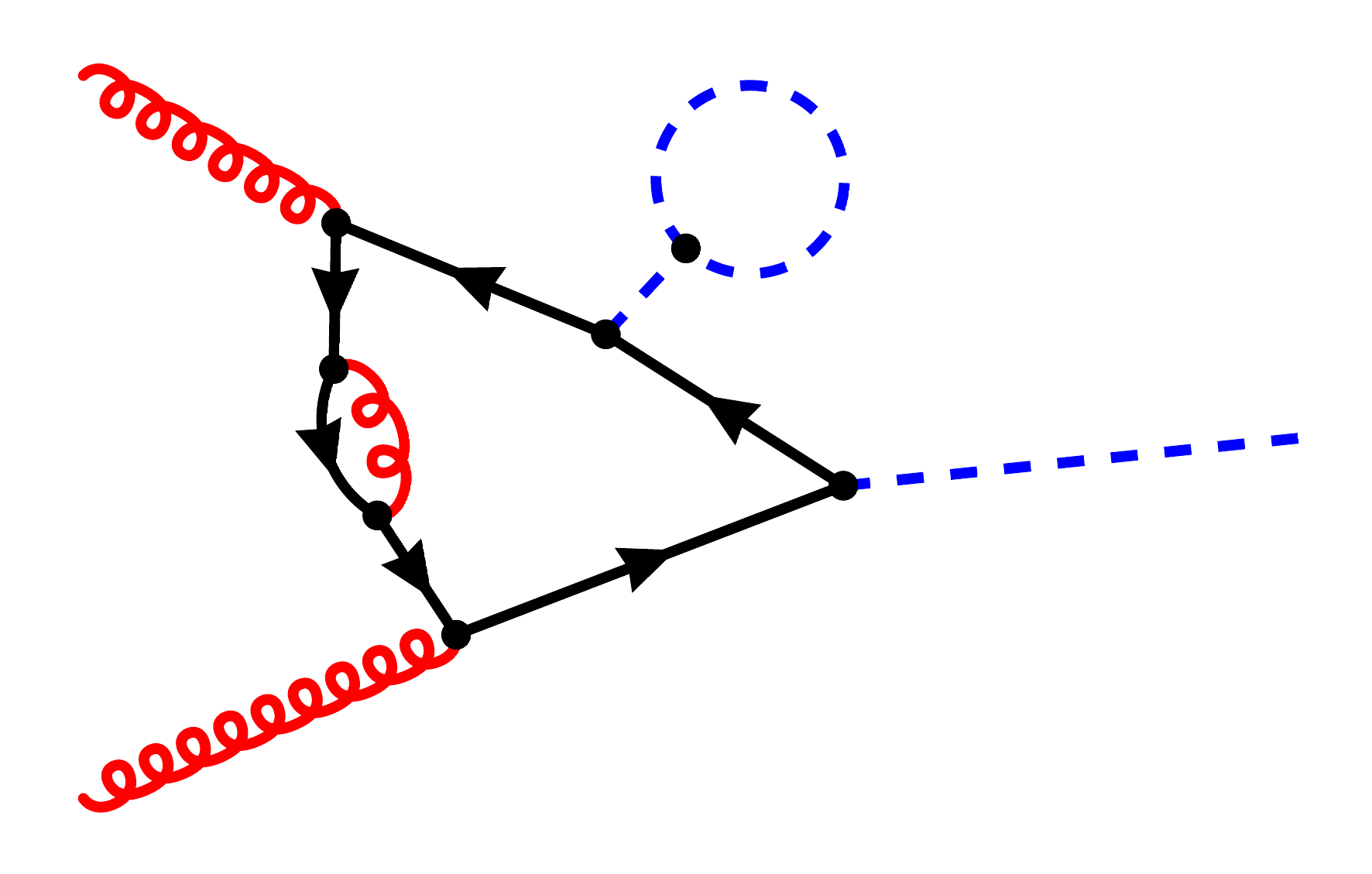} &
          \includegraphics[%
            width=.26\textwidth]%
                          {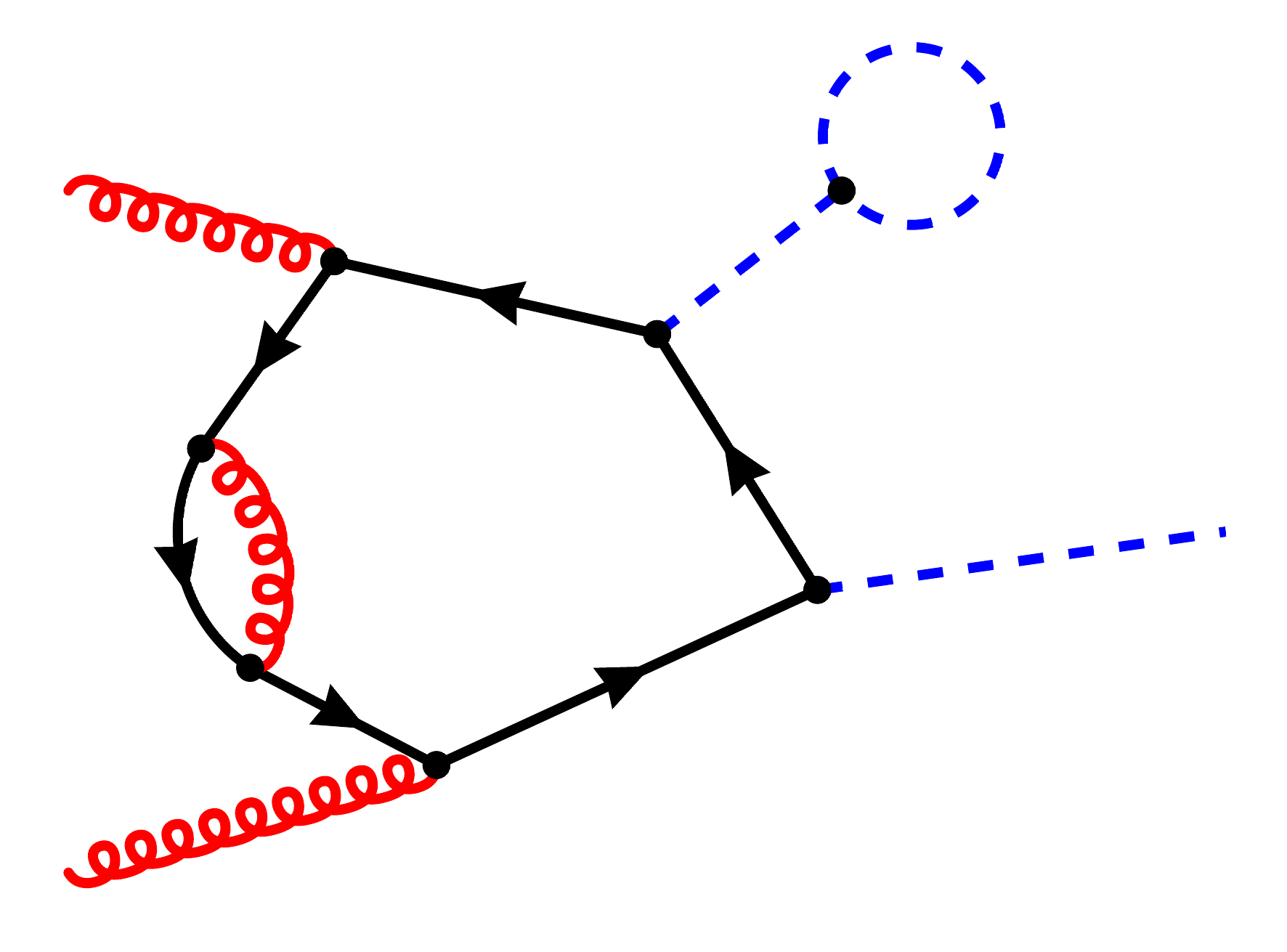} \\
                          (a) & (b)\\
          \includegraphics[%
            width=.26\textwidth]%
                          {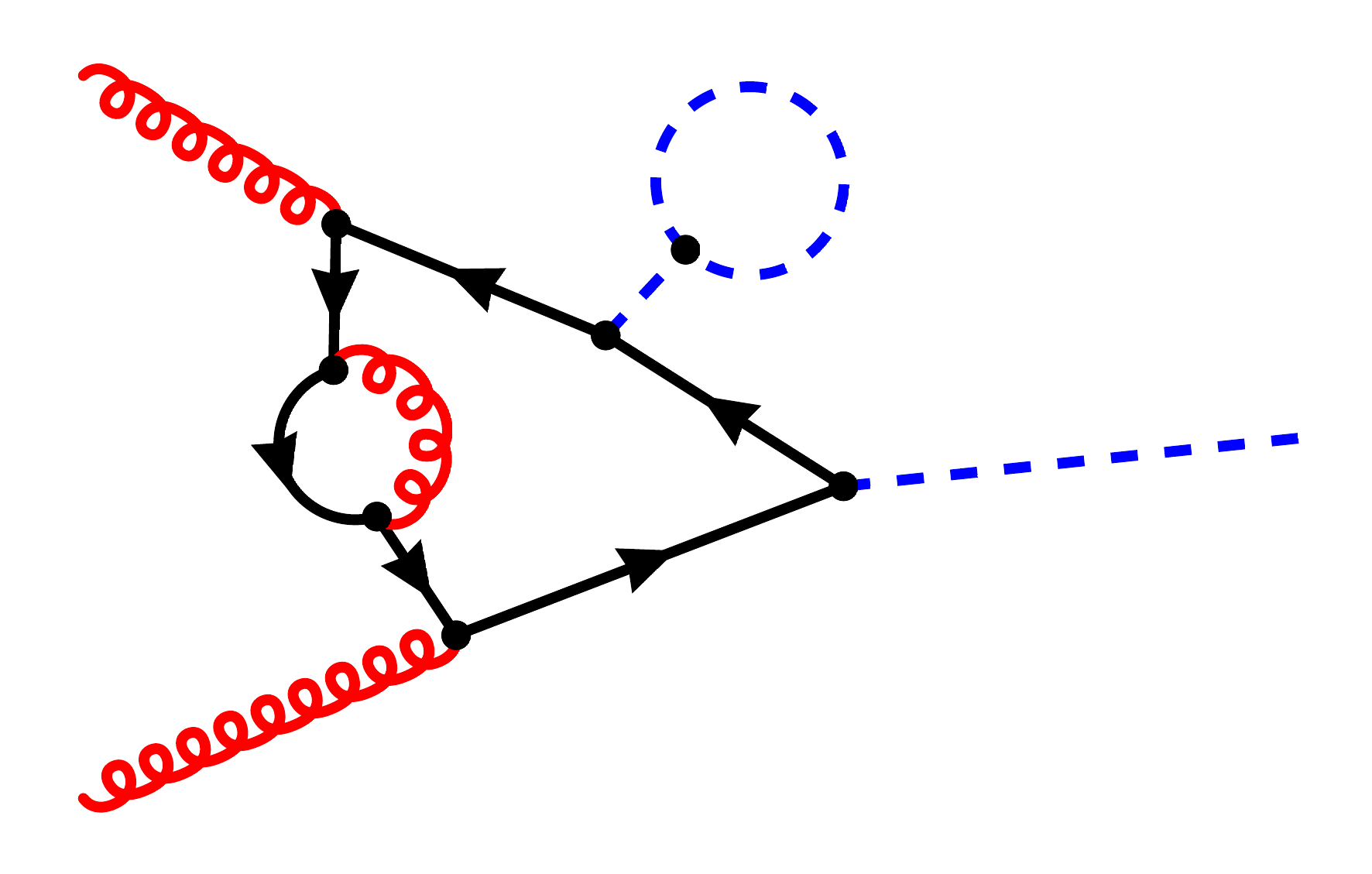} &
          \includegraphics[%
            width=.26\textwidth]%
                          {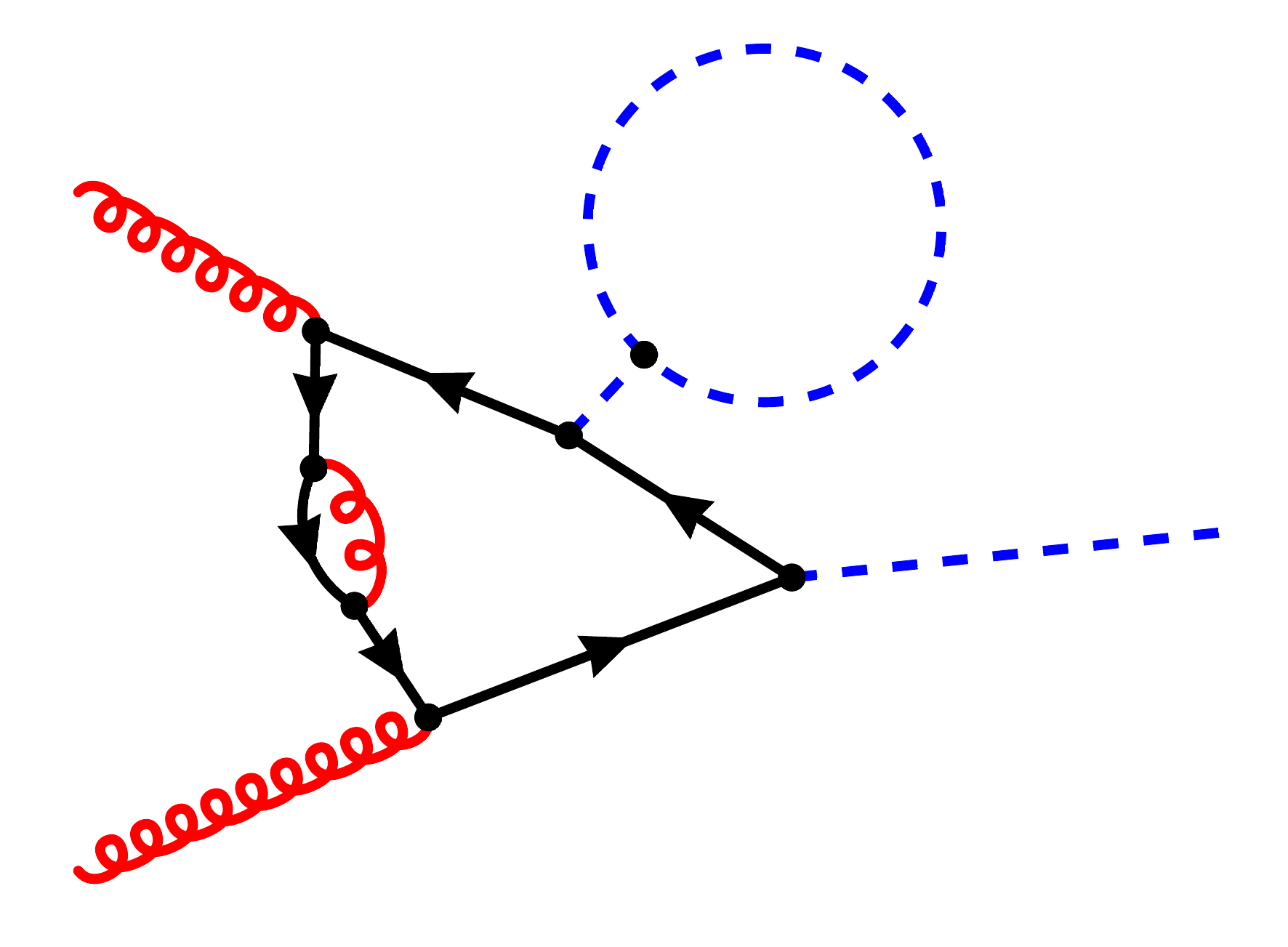}\\
                          (c) & (d)
    \end{tabular}
    \parbox{.9\textwidth}{
      \caption[]{\label{fig:spring_params}\sloppy Adjusting the layout
        parameters. Starting from some default setting in (a), the diagrams
        (b)--(d) visualize the effect of increasing the \textit{Length
          parameter}, the \textit{Multi-edge curve scale}, and the
        \textit{Self-loop radius scale}, respectively.  }}
  \end{center}
\end{figure}

We extended the algorithm to the multi-edge case, where the same two vertices
are connected by more than one line, and self-loops, where one line starts and
ends at the same vertex.\footnote{In particle physics, these are usually
referred to as ``self-energy'' and ``tadpole'' diagrams, respectively.} In
these cases, the lines will be drawn with a curvature, see
\cref{fig:spring_params}. Furthermore, external lines are interpreted as
either in- or outgoing, depending on where their open end is located on the
canvas. The drawing algorithm will preserve this character.

Most of the parameters for this algorithm can be modified by the user in order
to adjust the appearance of the diagrams to one's individual aesthetic
notion. They can be accessed via \menu[,]{File,Settings} which opens a new
window. From there, choose \menu[,]{Settings,Layout Options}.  Three of the
available options are visualized in \cref{fig:spring_params}.

\subsection{Manual application of the spring layout}\label{sec:manual_layout}

\begin{figure}
  \begin{center}
    \begin{tabular}{ccc}
      \raisebox{0em}{%
          \includegraphics[%
            clip,width=.3\textwidth]%
                          {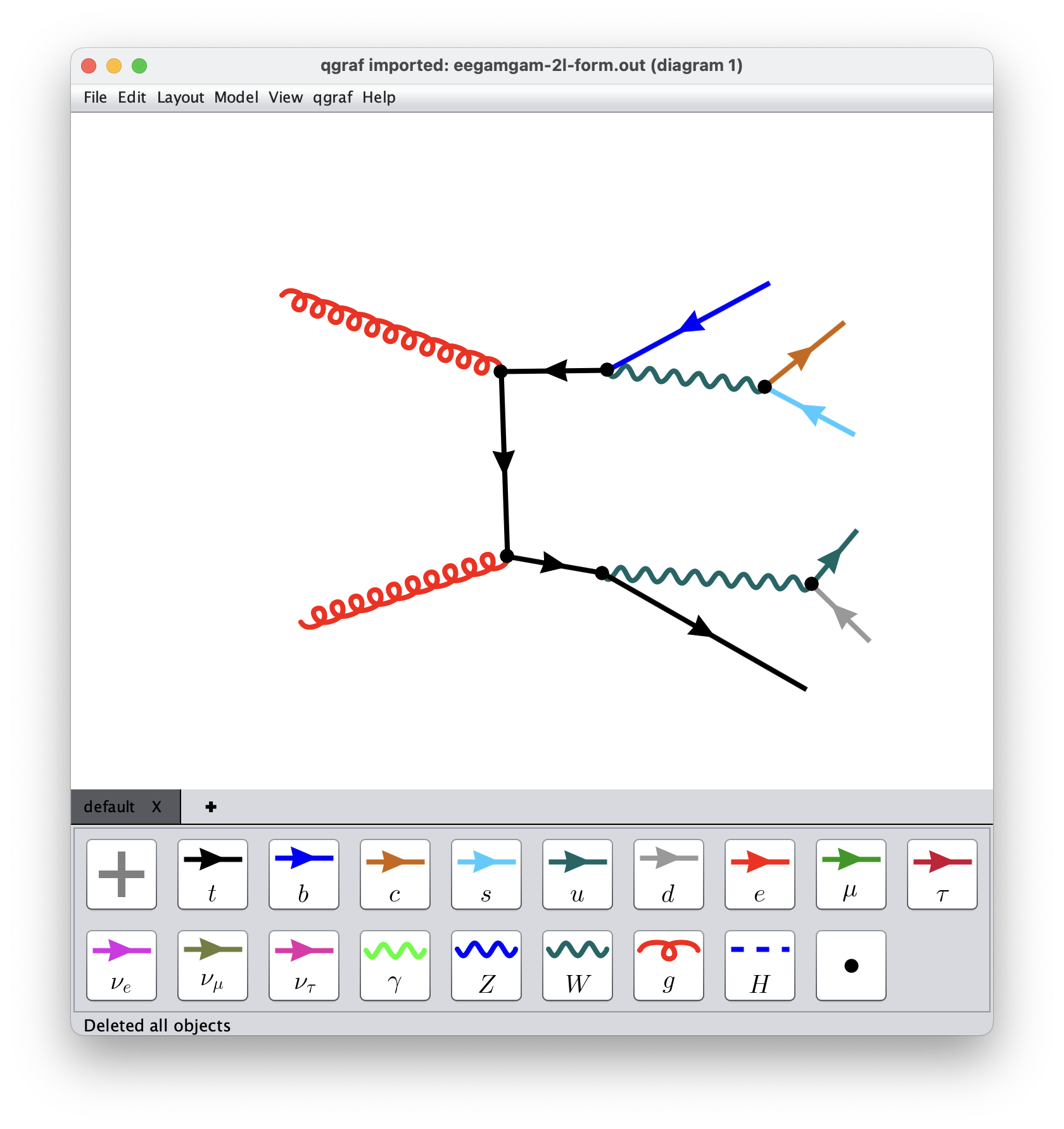}}
      &
      \raisebox{0em}{%
          \includegraphics[%
            clip,width=.3\textwidth]%
                          {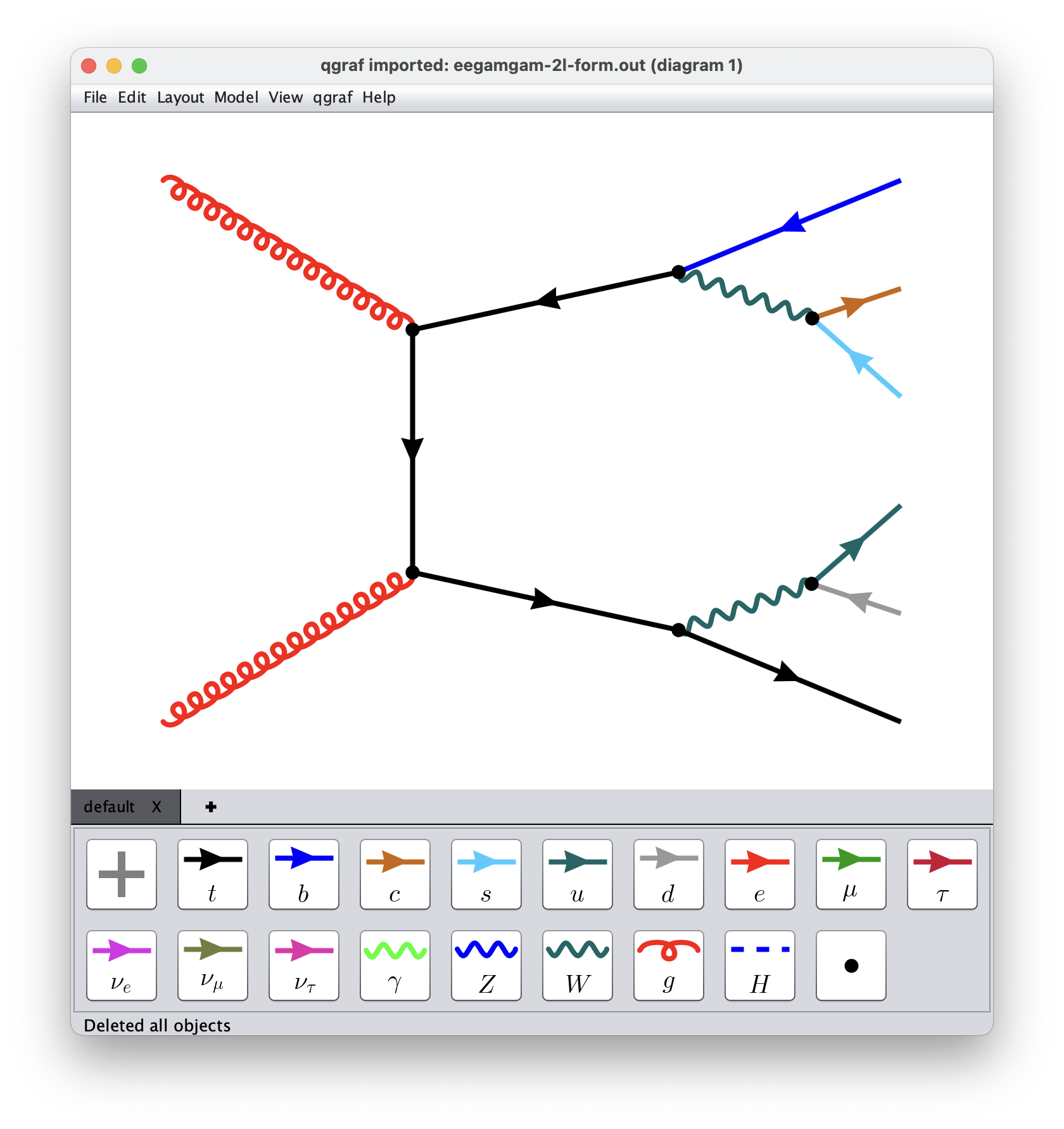}}
      &
      \raisebox{0em}{%
          \includegraphics[%
            clip,width=.3\textwidth]%
                          {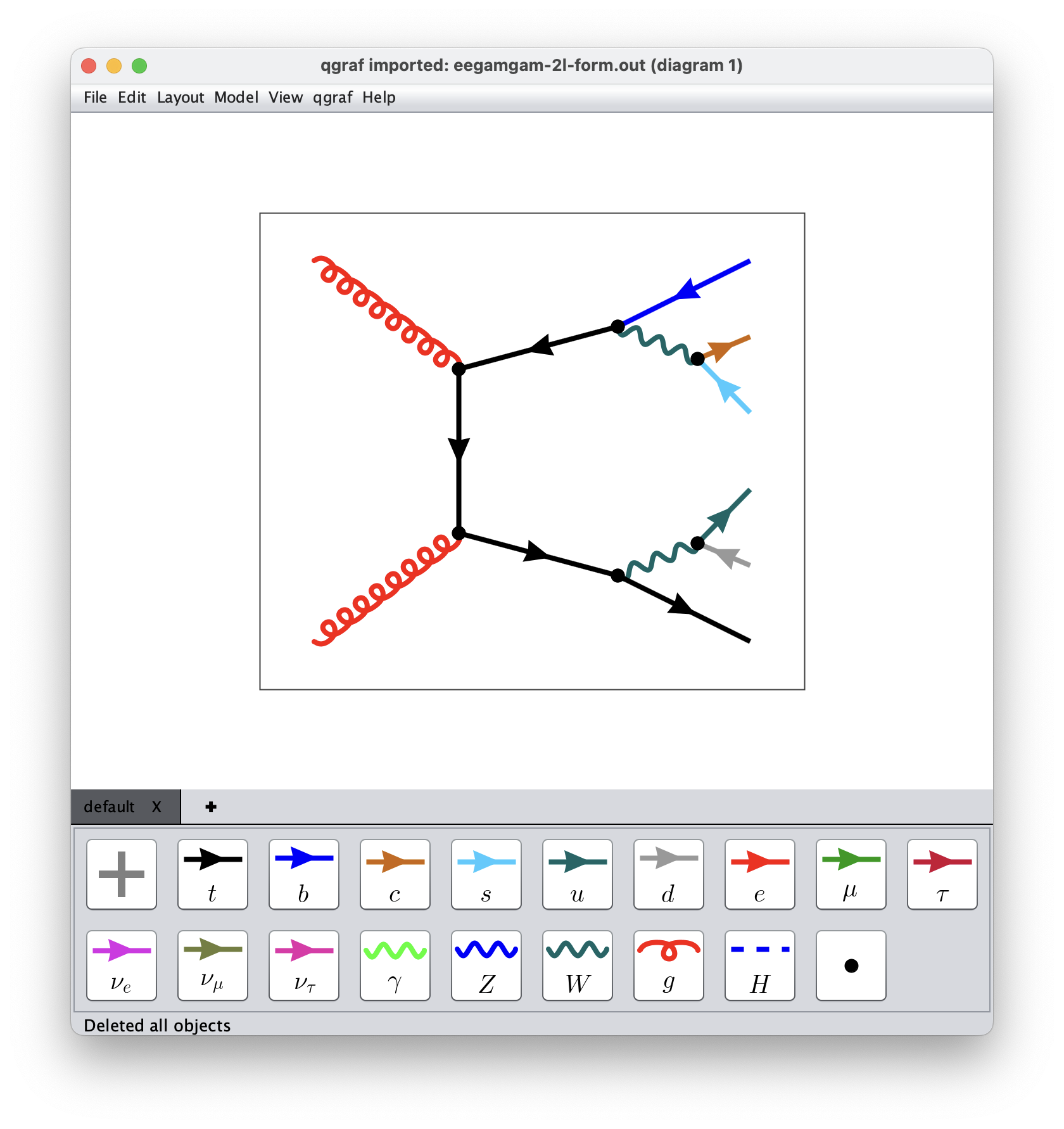}}\\
      (a) & (b) & (c)
    \end{tabular}
    \parbox{.9\textwidth}{
      \caption[]{\label{fig:auto-layout}\sloppy (a)~Rough drawing of a
        Feynman diagram in \feyngame; (b)~after application of the spring
        layout; (c)~spring layout with a manual bounding box.
    }}
  \end{center}
\end{figure}

The spring layout functionality of \feyngame{} is not restricted to Feynman
graphs produced by \qgraf. It may be applied to any diagram that the user
draws on the canvas. This makes drawing of individual Feynman diagrams even
more efficient than in previous versions of \feyngame. Imagine you would like
to draw a representative tree-level Feynman diagram for the \sm\ process
$gg\to t\bar t$ including the decay of the top quark to (quasi-)stable
particles. A ``raw'' version drawn with \feyngame{} is shown in
\cref{fig:auto-layout}\,(a). Due to the large number of particles involved, it
appears rather skew, and it would take some effort to make it more symmetric
manually. But \feyngame{-3.0} offers the option to turn any diagram into
spring layout. Choosing \menu[,]{Layout,Apply automatic layout} with this
diagram on the canvas will produce \cref{fig:auto-layout}\,(b), for
example. An easy option to modify the proportions of the diagram from their
default values is to define a bounding box via \menu[,]{Layout, Manual
  Bounding Box}, which allows one to arrive at \cref{fig:auto-layout}\,(c).

\subsection{Limitations and caveats}

\begin{figure}
  \begin{center}
    \begin{tabular}{ccc}
      \raisebox{0em}{%
          \includegraphics[%
            clip,width=.3\textwidth]%
                          {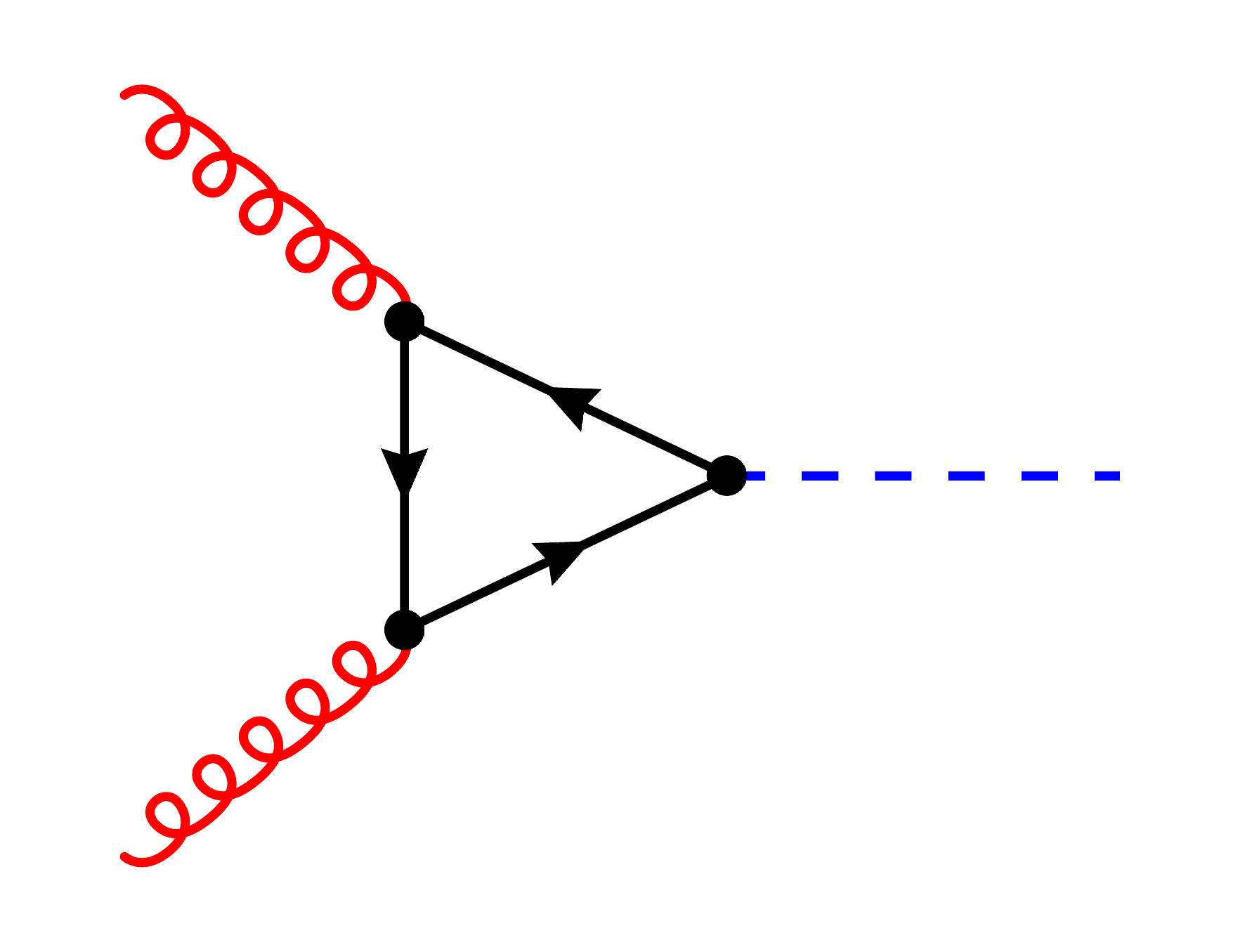}}
      &
      \raisebox{0em}{%
          \includegraphics[%
            clip,width=.3\textwidth]%
                          {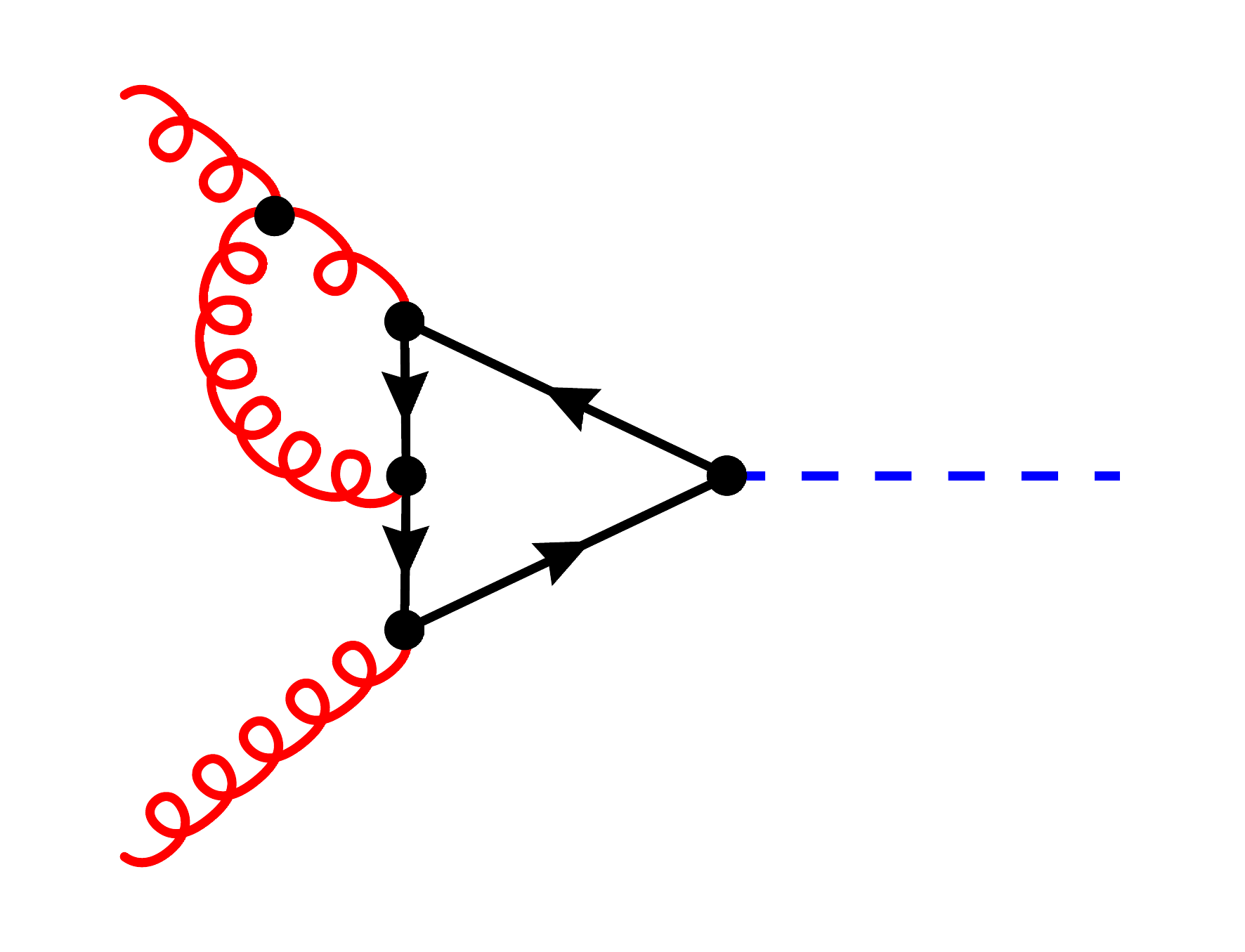}}
      &
      \raisebox{0em}{%
          \includegraphics[%
            clip,width=.3\textwidth]%
                          {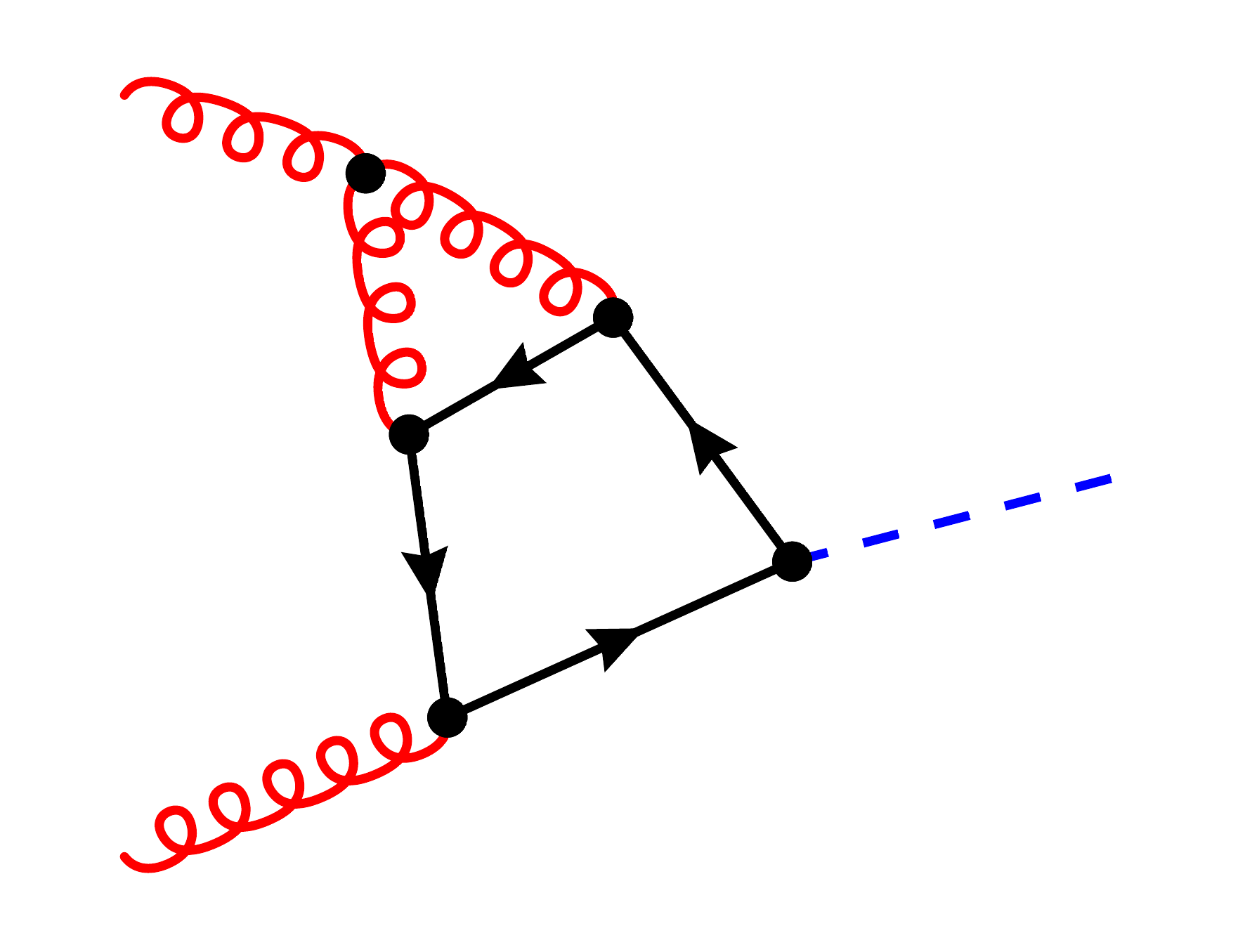}}\\
      (a) & (b) & (c)\\
    \end{tabular}
    \parbox{.9\textwidth}{
      \caption[]{\label{fig:triangle}\sloppy
        (a)~Leading order contribution to the process $gg\to H$ in
        spring layout; (b)~an \nlo\ contribution to this process as drawn by a
        practitioner; (c)~same contribution as in (b), but in the spring layout.
    }}
  \end{center}
\end{figure}

The spring layout typically manages to draw the diagram of an arbitrary graph
in a ``reasonable'' way. For example, a human can easily tell whether two
graphs have the same topology or not if both of them are visualized in the
spring layout. This does not mean that the diagrams produced by the spring
layout agree with the way a human would draw them, for several reasons. On the
one hand, every human has their own understanding of what is visually most
pleasing.  On the other hand, quite often practitioners will draw Feynman
diagrams from the point of view of perturbation theory. For example, given a
diagram that contributes to a certain process at leading order, a subset of
higher-order contributions is given by ``dressing'' this leading-order diagram
with additional lines. In this case, one often simply attaches the
``corrections'' to the leading-order diagram while keeping its shape
unchanged. Consider, for example, the diagram in
\cref{fig:triangle}\,(a)
which contributes to Higgs production in gluon fusion at \lo. A
\nlo\ \qcd\ contribution is obtained by simply attaching a gluon to this
diagram as done, for example, in \cref{fig:triangle}\,(b). Applying the spring layout,
however, results in \cref{fig:triangle}\,(c), and no researcher working in the
field would draw this \nlo\ contribution in this form, unless there is a very
specific reason for it.

Finally, the original spring layout algorithm relies on straight lines
only. In certain cases this leads to lines that are drawn on top of each
other. As mentioned above, we extended the spring layout to catch self-energy
loops and draw them with curved lines, but this is only a subset of cases
where overlays of lines can occur.

Naturally, all these limitations become more severe the larger the number of lines in the
diagram.

\section{Amplitudes}\label{sec:amplitude}

\begin{figure}
  \begin{center}
    \begin{tabular}{cc}
      \begin{minipage}[t]{.48\textwidth}
      \includegraphics[%
        width=1\textwidth]{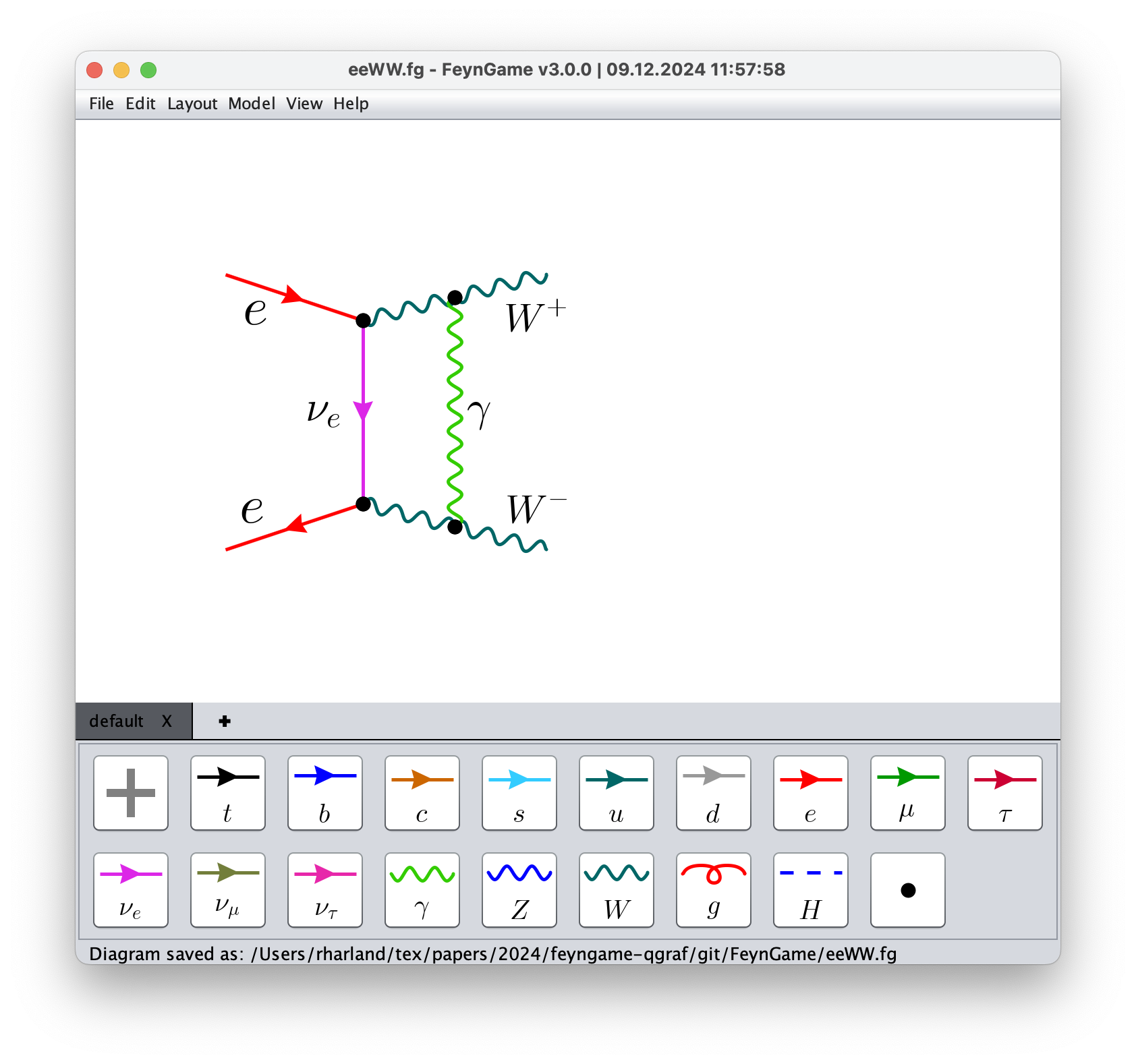}\\[-14em]
      \hspace*{9em}\includegraphics[%
        width=.5\textwidth]{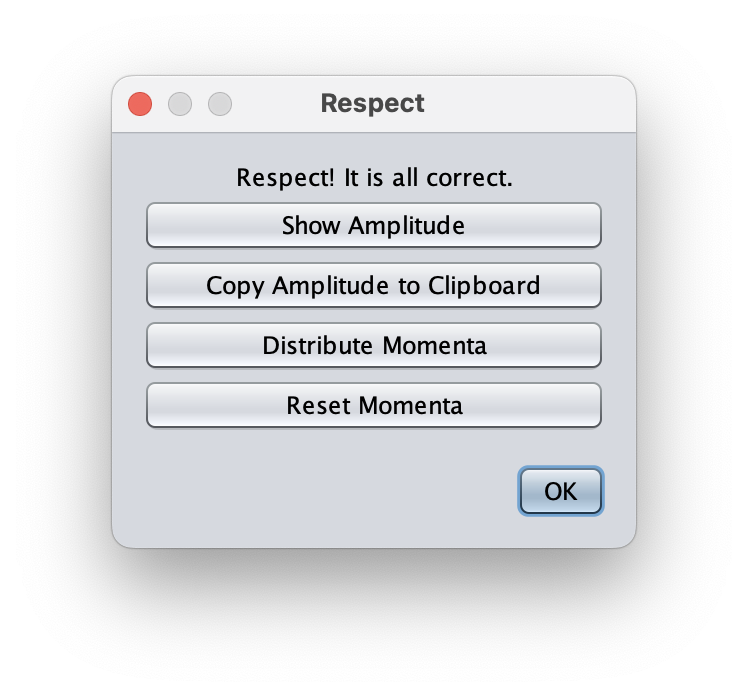}
      \end{minipage} &
      \begin{minipage}[t]{.48\textwidth}
      \includegraphics[%
        width=\textwidth]{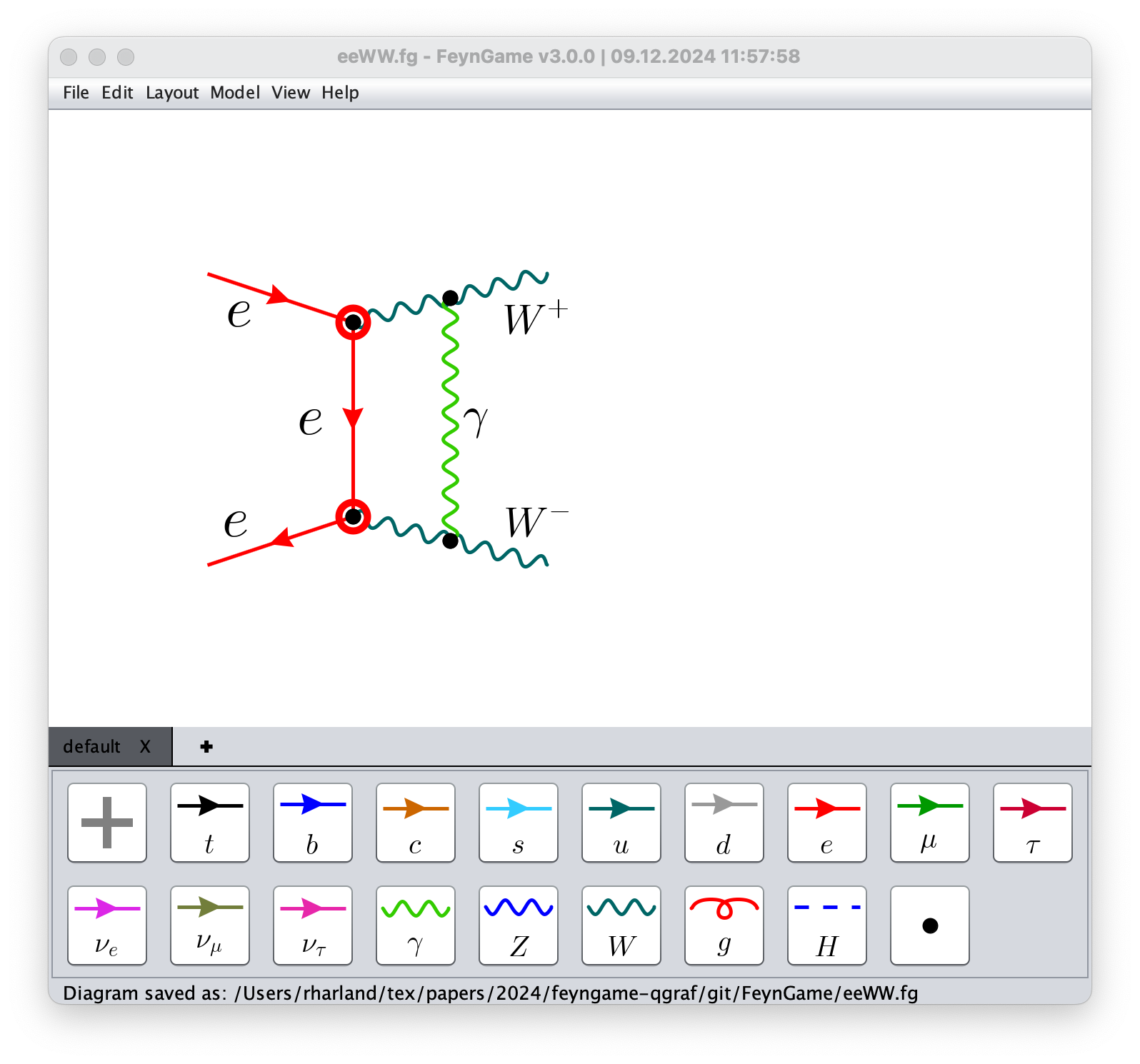}\\[-14em]
      \hspace*{9em}\includegraphics[%
        width=.5\textwidth]{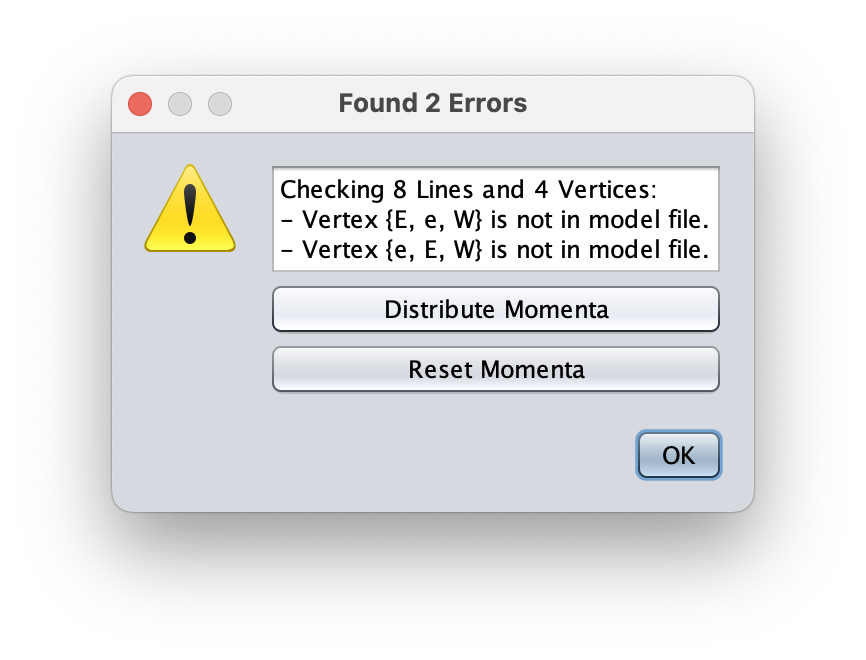}
      \end{minipage}\\
      (a) & (b)
    \end{tabular}
    \parbox{.9\textwidth}{
      \caption[]{\label{fig:f-error}\sloppy
        Checking the validity of a Feynman diagram. (a)~The diagram exists in
        the \sm; \feyngame\ then offers to produce the algebraic expression
        for the amplitude. (b)~The diagram is in conflict with the Feynman
        rules of the \sm; \feyngame\ reports the erroneous vertices and marks
        them by red circles.
    }}
  \end{center}
\end{figure}

Even though it was already introduced in \feyngame{-2.1} and briefly
discussed in \citere{Harlander:2024qbn}, let us describe the feature of
generating the mathematical expression for an amplitude with \feyngame{} in a
bit more detail at this point.

Assume that we work with \feyngame{}'s default model, corresponding to the
\sm. After drawing a diagram, one can ask \feyngame{} to check whether the
diagram is consistent with the Feynman rules by pressing \keys{f} or
\menu[,]{View,Check for errors}.  If the diagram is valid, \feyngame{} will
offer to generate the corresponding Feynman amplitude, or to distribute
momenta among the lines of the diagram, preserving momentum conservation (see
\cref{fig:f-error}\,(a)). The amplitude is generated in \LaTeX{} format and
can be pasted immediately into a \LaTeX\ document (line breaks need to be
taken care of manually).

On the other hand, if the diagram involves vertices that are not part of the
model, \feyngame{} will report the erroneous vertices and mark them by red
circles in the canvas, see \cref{fig:f-error}\,(b). Obviously,
\feyngame\ cannot produce an amplitude in this case. However, it still offers
to distribute momenta among the lines of the diagram.

The basic functionality for producing the amplitude is not restricted to
\LaTeX{} output, of course. To illustrate this, let us look at the full
definition of the top quark propagator in the default model file:
\begin{lstlisting}[%
    style=qgraf,
    caption={Definition of the top-quark line in the default
      \feyngame\ model file.},
    label=lst:top-propagator
  ]
  [t, T, PLAIN,
    color=ff000000, stroke=4.0, dash=false, double=false,
    showArrow=true, arrowSize=21, arrowPlace=0.5, arrowDent=2,
    arrowHeight=18,
    label=t,
    indices=<0|1|0|1>,
    propagator=
    \frac{i\delta_{<idxE.cf><idxS.cf>}(<p.slash>+<m>)_{<idxE.di><idxS.di>}}
         {<p>^2-<m>^2+i\epsilon},
    external=u_{<idx.di>}^{<idx.cf>}(<p.vector>) |
             \bar{u}_{<idx.di>}^{<idx.cf>}(<p.vector>),
    external*=\bar{v}_{<idx.di>}^{<idx.cf>}(<p.vector>) |
              v_{<idx.di>}^{<idx.cf>}(<p.vector>),
    anticommuting=true]
\end{lstlisting}
The first few entries determine the graphical appearance of the line within
\feyngame{}, see also \cref{sec:sync}. Starting from \code{indices=<0|1|0|1>},
however, the entries define the Feynman rule for this line. Specifically,
this entry defines the number of the Lorentz, Dirac, adjoint, and fundamental
color indices carried by this object.\footnote{This list can be
arbitrarily extended by additional indices.} For the remaining entries, their
notation should be sufficiently suggestive for an expert reader to derive
their meaning. We refrain from explaining them here in detail, but rather
refer the reader to the template file \texttt{examples/amplitude.model} which
is part of the \feyngame\ distribution and contains a detailed documentation.

Despite the fact that these entries are specific for \LaTeX\ output, they
contain all information to generate the amplitude in other formats as well. In
particular, one may convert the entries to
\textit{\abbrev{FORM}}\,\cite{Vermaseren:2000nd,Kuipers:2012rf,Ruijl:2017dtg}
or \textit{Mathematica}~\cite{Mathematica} format, which allows the user to
directly continue with the algebraic or numerical evaluation of the
amplitude. Let us emphasize that this is not the main purpose of
\feyngame\ though; the calculation of Feynman amplitudes can be done much more
efficiently with other tools. But we hope that the amplitude feature of
\feyngame\ may be still be helpful, for example in an educational context.

\section{Further improvements}\label{sec:improvements}

In this section, we will discuss a number of new features in \feyngame{} which
have been implemented to facilitate its use. Some of these features were
already available in \feyngame{-2.1}, but we think it is worth including them
in this list.

\begin{figure}
  \begin{center}
    \begin{tabular}{cc}
      \raisebox{3em}{%
          \includegraphics[%
            clip,width=.4\textwidth]%
                          {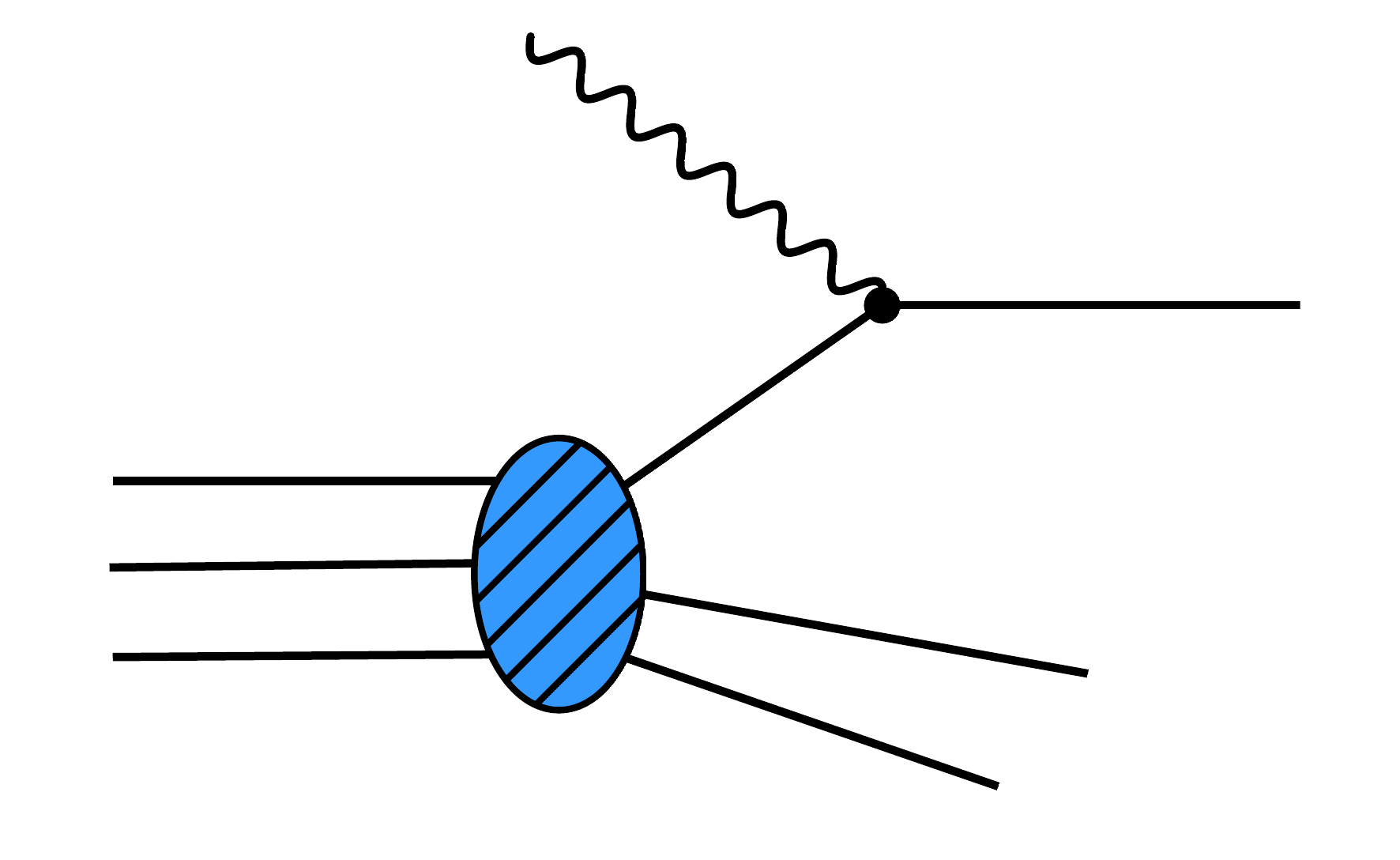}}
      &
      \raisebox{0em}{%
          \includegraphics[%
            clip,width=.4\textwidth]%
                          {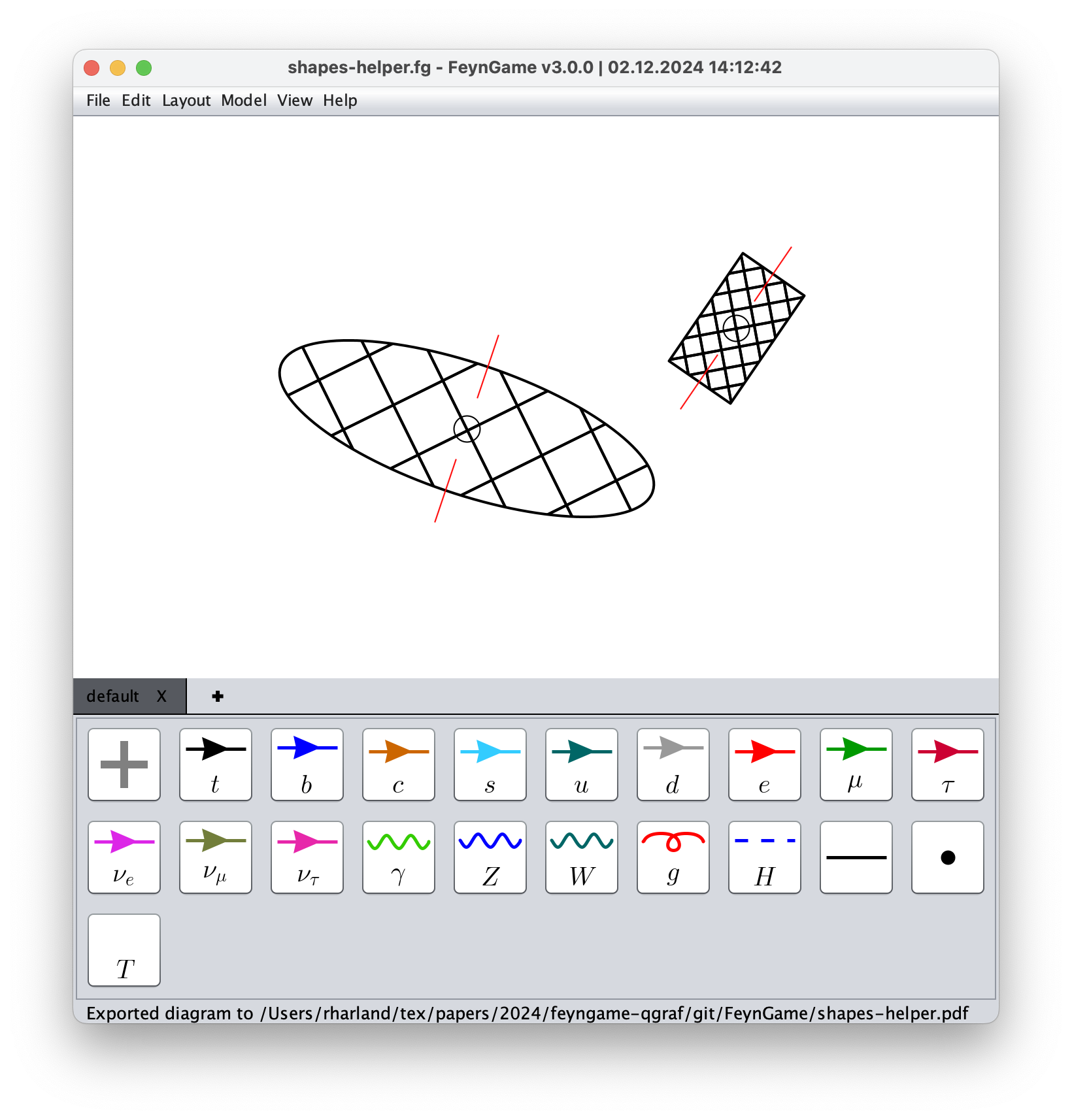}}
      \\
      (a) & (b)
    \end{tabular}
    \parbox{.9\textwidth}{
      \caption[]{\label{fig:shapes}\sloppy
        Shapes in \feyngame. (a)~Using a shape in a Feynman diagram;
        (b)~shapes on the canvas with \textit{helper lines} activated; this
        facilitates rotating and moving the shapes by mouse.
    }}
  \end{center}
\end{figure}

\begin{figure}
  \begin{center}
    \begin{tabular}{c}
      \raisebox{0em}{%
        \mbox{%
          \includegraphics[%
            width=.9\textwidth]%
                          {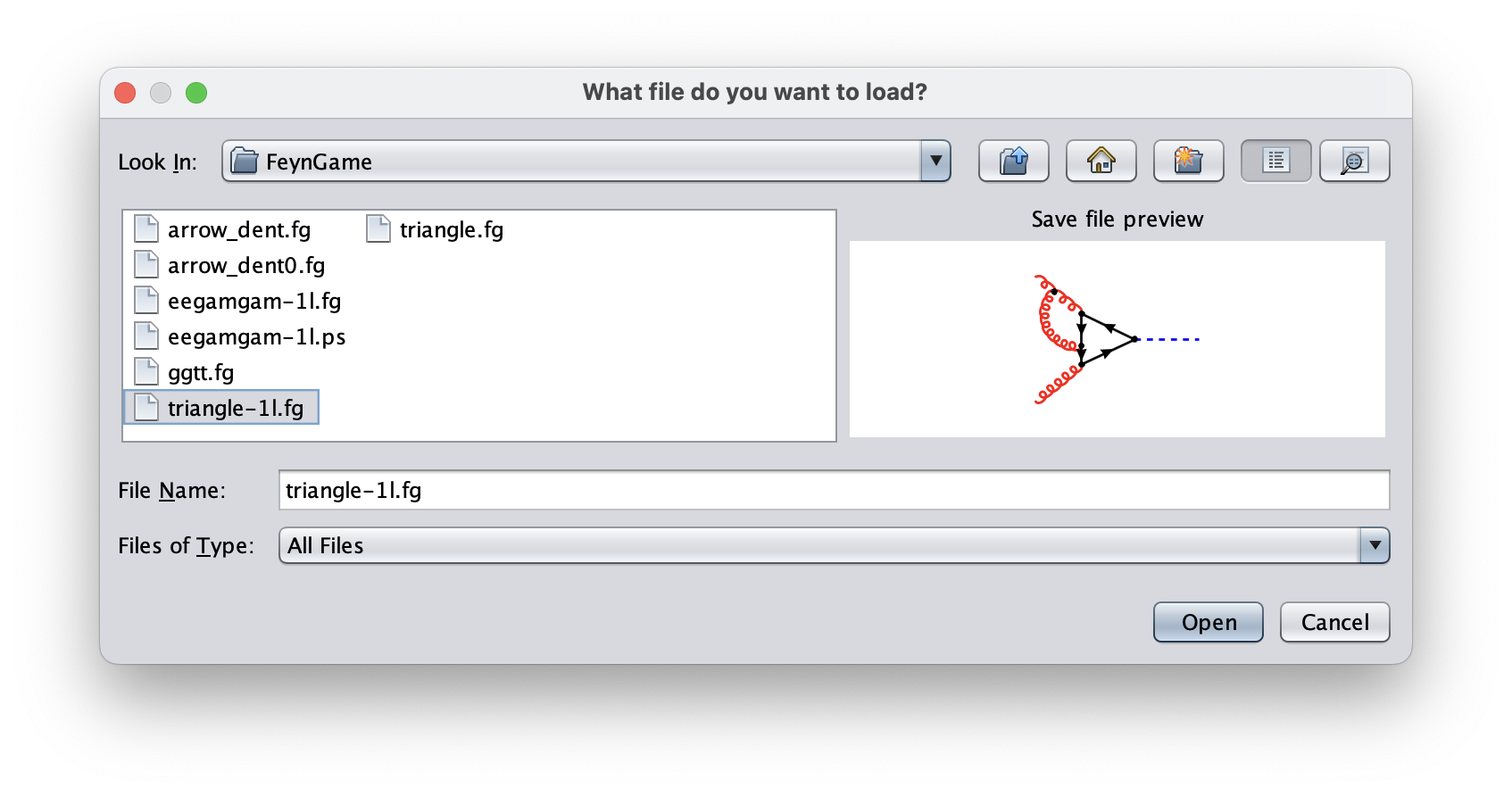}}}
    \end{tabular}
    \parbox{.9\textwidth}{
      \caption[]{\label{fig:chooser}\sloppy File chooser of \feyngame\ with a
        preview of the diagram.  }}
  \end{center}
\end{figure}

\begin{figure}
  \begin{center}
    \begin{tabular}{cc}
          \includegraphics[%
            width=.4\textwidth]%
                          {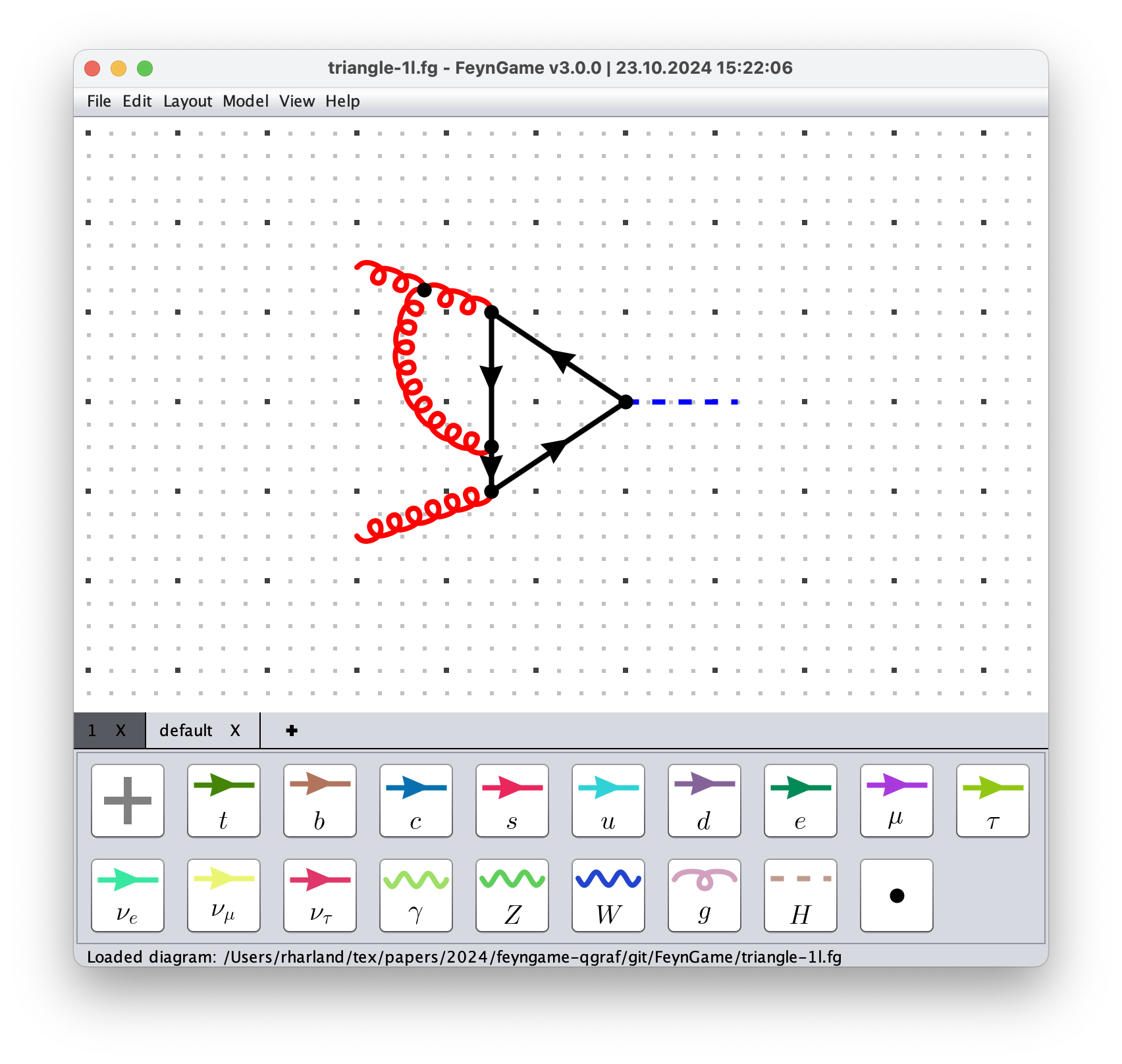} &
          \includegraphics[%
            width=.4\textwidth]%
                          {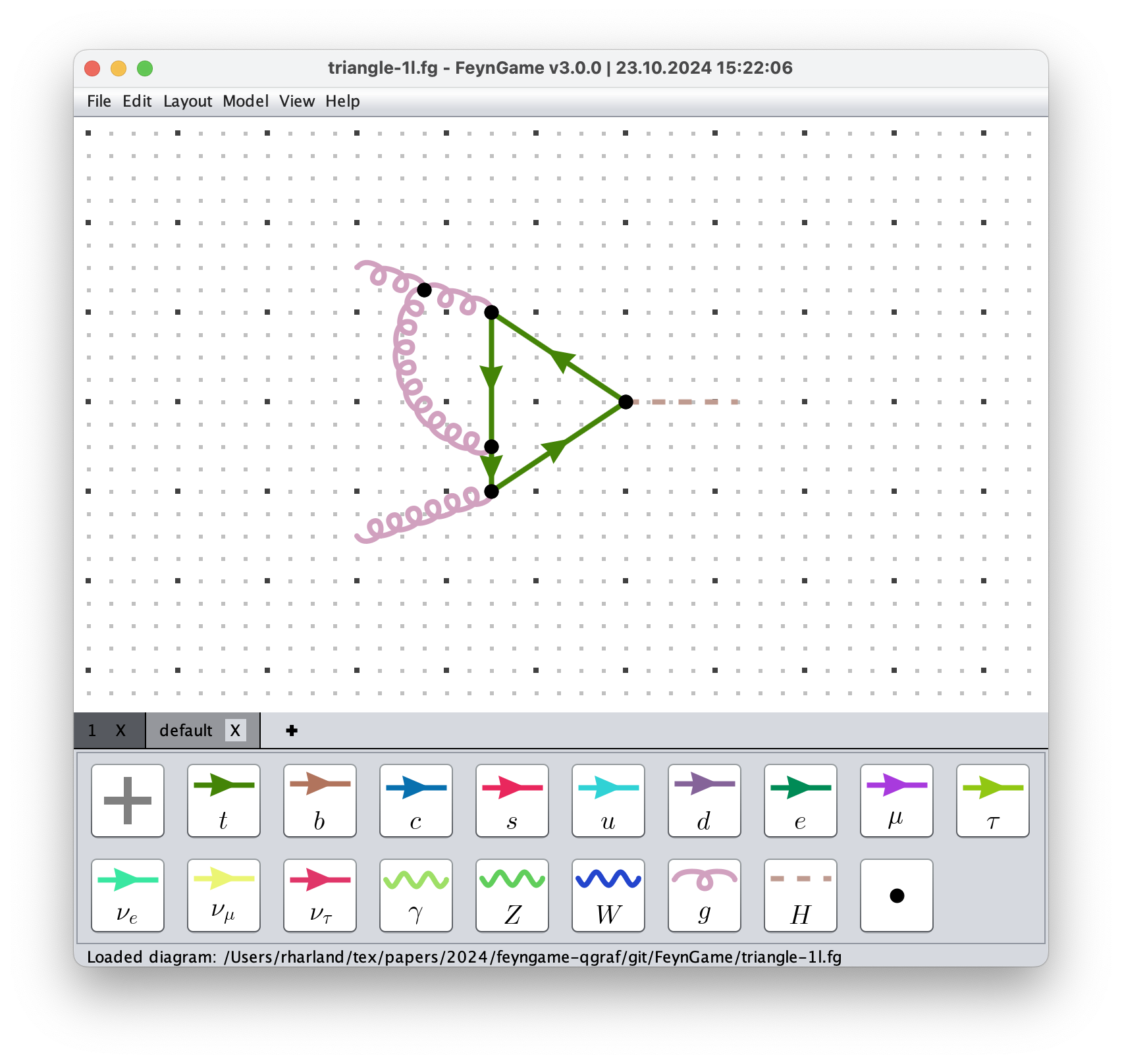}\\
                          (a) & (b)
    \end{tabular}
    \parbox{.9\textwidth}{
      \caption[]{\label{fig:apply_model}\sloppy (a) Diagram drawn with the
        default model; (b)~diagram after applying the current model.  }}
  \end{center}
\end{figure}

\begin{figure}
  \begin{center}
    \begin{tabular}[b]{cc}
    \begin{tabular}[b]{c}
      \raisebox{0em}{%
          \includegraphics[%
            width=.15\textwidth]%
                          {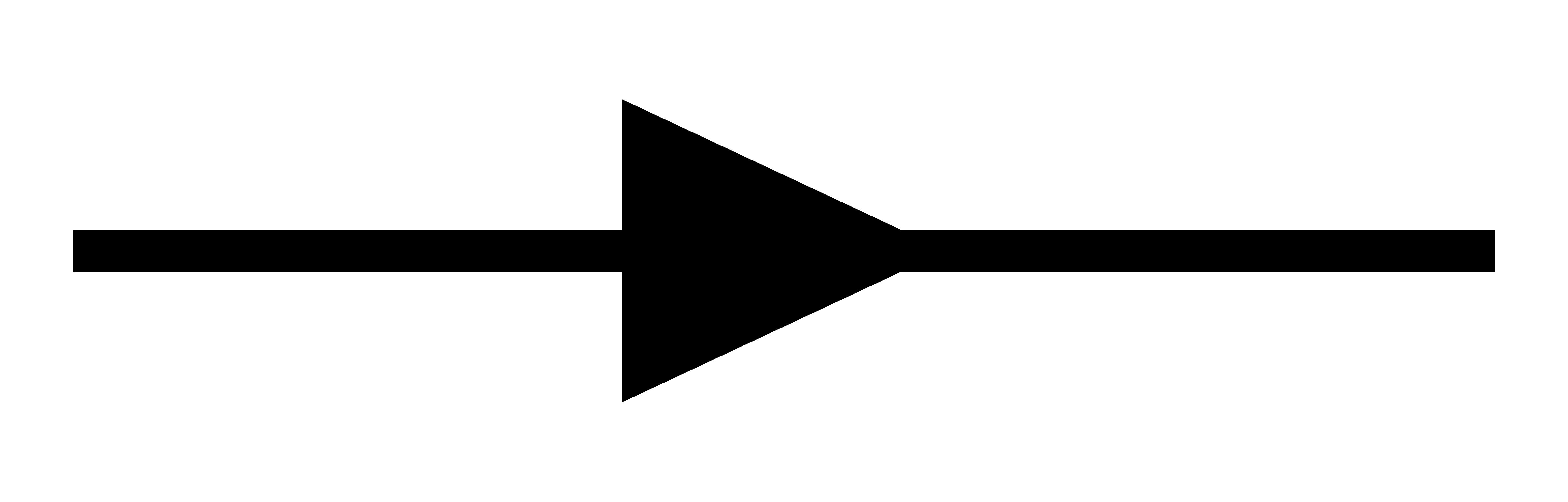}}\\
      \raisebox{0em}{%
          \includegraphics[%
            width=.15\textwidth]%
                          {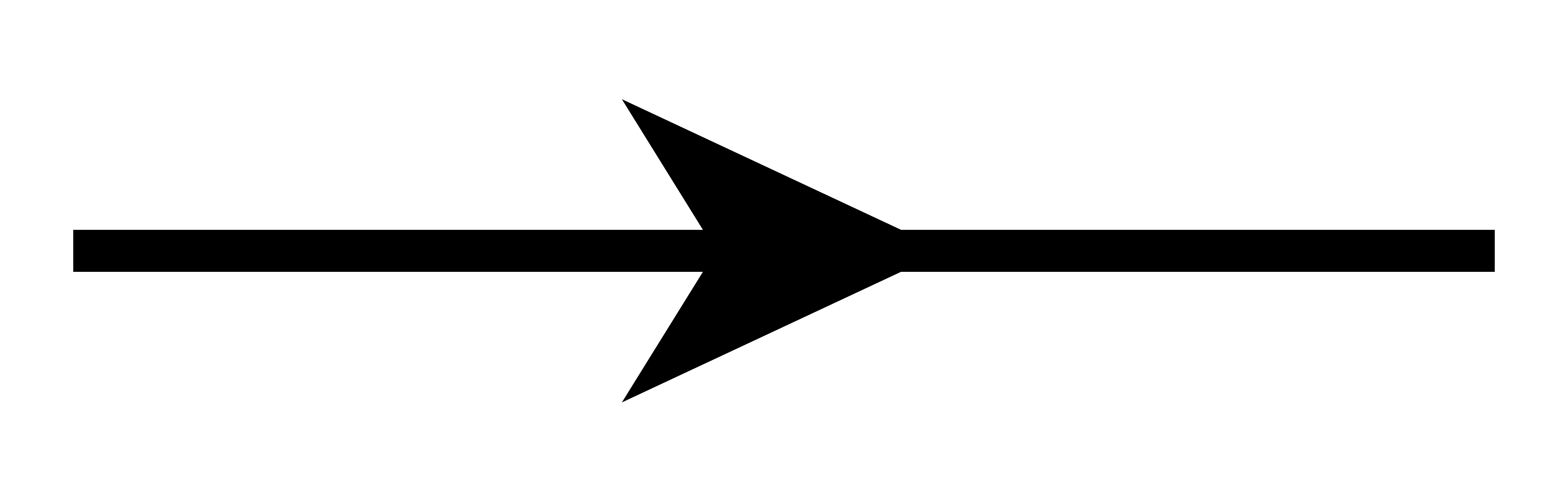}}\\
      (a)
    \end{tabular}
    &
    \begin{tabular}[b]{c}
      \raisebox{0em}{%
          \includegraphics[%
            width=.3\textwidth]%
                          {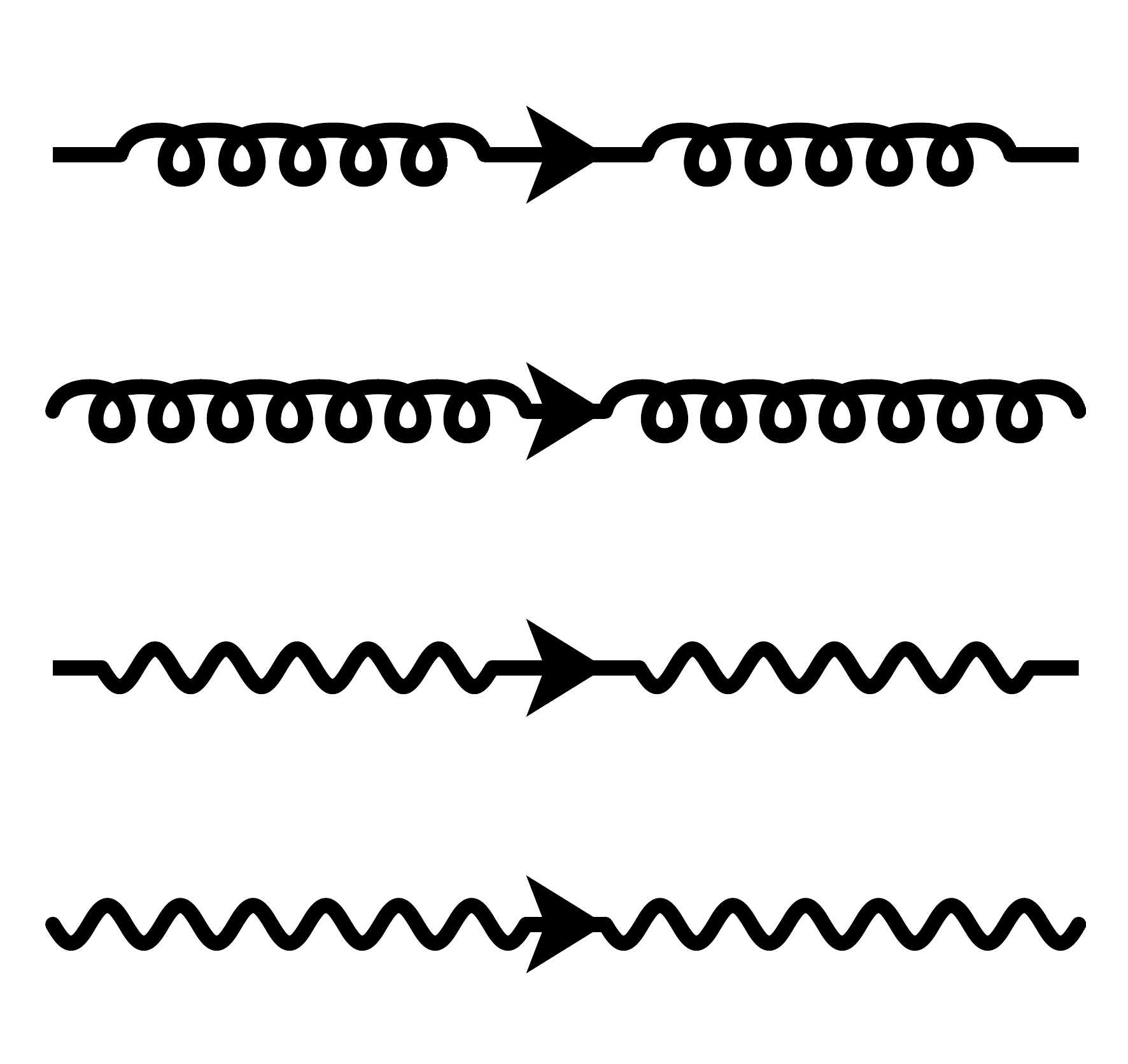}}
      \\
      (b)
    \end{tabular}
    \end{tabular}
    \parbox{.9\textwidth}{
      \caption[]{\label{fig:features}\sloppy
        (a)~Lines with arrows with parameter \texttt{arrowDent} equal to zero
        (upper) and non-zero (lower); (b)~spring and wave lines with arrows,
        and different settings for the parameters \texttt{straightLength} and
        \texttt{straightLengthArrow}.
    }}
  \end{center}
\end{figure}

\begin{description}
\item[Shapes:] Using \menu[,]{Edit,Add new shape}, one can paste a
  \textit{shape} onto the \feyngame\ canvas. Currently, ovals and rectangles
  are supported shapes, see \cref{fig:shapes}\,(a). For the manipulation of
  shapes, it is recommended to switch on the \textit{helper
    lines}~\cite{Harlander:2020cyh} by pressing \keys{h}, see
  \cref{fig:shapes}\,(b).
\item[File chooser preview:] \feyngame's file chooser now provides a preview
  of the selected diagram. This is very convenient for finding existing
  Feynman diagrams in a directory. For example, \cref{fig:chooser} shows the
  file chooser that opens with \menu[,]{File,Open} in the directory that
  contains the \fgfile{}s related to this paper.
\item[\hypertarget{pos:convert}{Conversion} to a different model:] A \fgfile{}
  which was drawn with a foreign model file can be converted to the current
  model by loading the \fgfile{} into the \feyngame{} window and
  selecting \menu[,]{Model,Apply model to canvas}. This feature can also be
  accessed from the command line:
\begin{lstlisting}[style=shell,
    label={lst:apply-model}]
  $ feyngame --diagram <path_to_diagram.fg> --model <new_model> --apply-model --export <diagram>
\end{lstlisting}
  This command will produce a \abbrev{PDF} file \code{<diagram>.pdf} of the
  diagram, using the \texttt{<new\_model>} file. See also
  \cref{sec:commandline}.
\item[Improved grid:] The grid of \feyngame{-3.0} (toggled with \keys{g}, or
  \menu[,]{View,Grid}) consists of two overlayed grids of different
  periodicity, marked by different grid dots (see, e.g.,
  \cref{fig:apply_model}).  This significantly facilitates the symmetric
  positioning of lines and vertices etc.
\item[Bounding box:] The user may set a bounding box on the canvas by using
  the menu items
  \menu[,]{Layout,Show bounding box}. It can be adjusted either by mouse, or
  via \menu[,]{Layout,Configure bounding box}. When exporting the diagram to
  some image format, \feyngame{} will restrict the exported area to the
  bounding box. The bounding box will be saved with the diagram in the
  \fgfile. It can also be re-used for the next diagram to be drawn on the
  canvas. This enables one to draw a series of diagrams, all of whose
  \pdf\ files will have the same proportions.
\item[Improved export options:] Like in previous versions, the diagrams can be
  exported to all common image formats. For presentations, the format of
  choice is probably \abbrev{PNG} which provides the option of a transparent
  background. The most convenient way when preparing a Keynote or PowerPoint
  presentation is to copy the image to the clipboard (either by using
  \keys{\cmd-i}, or \menu[,]{File,Export to clipboard}) and pasting it into
  the presentation.  For publications, one may prefer vector graphics like
  (Encapsulated) PostScript or \abbrev{PDF}.\footnote{The latter will be
  produced via the detour of first producing PostScript, and subsequently
  converting it to \abbrev{PDF} using \texttt{ps2pdf} (if available on your
  system). }
\item[Arrows:] They now have the additional parameter \code{arrowDent}. Its
  effect should be clear when looking at \cref{fig:features}. Additionally,
  the height and length of arrows can now be adjusted independently from one
  another.
\item[Spiral and wave lines:] Several improvements have been made for these
  lines, see \cref{fig:features}:
  \begin{itemize}
  \item There is a straight section at both of their ends. Its (absolute) length
    is set by the parameter \code{straightLength} in the model file, and can
    be adjusted in the \editframe.
  \item If the \code{arrow} attribute is switched on, the arrow is drawn at a
    position where a wiggle crosses the longitudinal axis. One may also add
    straight sections before and after the arrow whose length is set by the
    parameter \code{straightLengthArrow} in the model file and can be set in
    \editframe.
  \end{itemize}
\item[Fermion direction:] The direction of fermion lines has been adjusted to
  be in accordance with \qgraf.
\item[\hypertarget{pos:action}{Action messages:}] \feyngame{-3.0} reports its
  last action at the bottom of the \feyngame{} window. This is helpful, for
  example, when saving or exporting diagrams, in which case \feyngame{} will
  report on the exact location of the saved or exported file.
\item[Log file:] Some information about \feyngame{}'s internal actions are
  logged into a file which the user can view via \menu[,]{Help,View Log}.
\item[Check for updates:] When connected to the internet, \feyngame{} will
  check for updates upon startup and inform the user about it.
\end{description}

\section{Installation and modes of operation}\label{sec:commandline}

\subsection{Download, installation, and graphical mode}

Compiled versions of \feyngame\ can be
obtained from
\begin{center}
  \url{https://web.physik.rwth-aachen.de/user/harlander/software/feyngame}
\end{center}
These include a \code{jar} file which can be run from a terminal as
\begin{lstlisting}[style=shell]
  $ java -jar FeynGame.jar
\end{lstlisting}
Alternatively, we provide the executable \code{feyngame} which can be opened
simply by saying
\begin{lstlisting}[style=shell]
  $ ./feyngame
\end{lstlisting}
in a terminal window. For MacOS, we also provide \code{FeynGame.dmg}, which
contains an app that can be moved into the \code{/Applications} folder and
linked to the \textit{Dock}, for example.
The source code of \feyngame\ is located at
\begin{center}
  \url{https://gitlab.com/feyngame/FeynGame}
\end{center}
Note that, in any case, you need to have a working version of \textit{Java}
installed on your computer, ideally version~9 or newer.

By default, the commands above (or opening the app) will open a dialogue
window which allows one to choose between the \textit{drawing mode} and the
\textit{game mode} of \feyngame. The latter is described in detail in
\citere{Harlander:2020cyh}. In both cases, \feyngame\ will subsequently open
its main window. However, \feyngame\ provides a number of options for the
calls from a terminal. Assuming that the \code{feyngame} executable has been
placed into the user's path, the full list can be viewed via
\begin{lstlisting}[style=shell,
    label={lst:help}]
  $ feyngame --help
\end{lstlisting}

\subsection{Command-line mode}

Many of \feyngame{}'s functionalities can be accessed via its command-line
mode, i.e.\ without opening its \gui. This may be particularly helpful in
combination with the new \qgraf\ import functionality, see
\cref{sec:qgraf}. An example for a command-line operation of \feyngame\ could
look as follows:
\begin{lstlisting}[style=shell,
    label={lst:export-all}]
  $ feyngame --qgraf-file <qgraf-out> --qgraf-style form.sty --export all.pdf --diagrams-per-page 6 6
\end{lstlisting}
This command reads the \qgraf\ output file \texttt{<qgraf-out>} which we
assume to be produced with the \qgraf\ style file \texttt{form.sty}, and
exports the diagrams in \pdf\ format to the file \texttt{all.pdf}, putting six
diagrams per row and column (similar to \cref{fig:eegamgam-2l-all}). In this
case, \feyngame{} effectively acts as a converter from \qgraf{} output to
\pdf{}. This is done without ever showing the \feyngame{} window and can be
included in automated scripts as part of a larger workflow\footnote{This
conversion internally still uses a hidden window, so it will not work in a
headless graphics environment.}.

Another application of the command line mode is to re-export diagrams to, say,
\pdf\ format after modifying the model file, as described in
\cref{sec:improvements}, \hyperlink{pos:convert}{\textit{Conversion to
    different model}}.

\section{Conclusions and outlook}\label{sec:conclusions}

We have presented version~3.0 of \feyngame\ and described its central new
features. Among them is the capability to display Feynman graphs generated by
\qgraf, and to produce a mathematical expression for the amplitude of an
arbitary \sm\ Feynman diagram. A number of other improvements are included,
such as a double grid and the preview of diagrams in the file chooser.

Future projects within \feyngame\ include improved visual representation of
crossed lines, updates for \feyngame's game mode, and a tutorial on Feynman
diagrams for high-school and graduate students. 

\paragraph{Acknowledgments.}
This research was supported by the \textit{Deutsche Forschungsgemeinschaft
  (DFG, German Research Foundation) under grant 396021762 - TRR 257}.

\bibliographystyle{utphys}
\bibliography{paper}

\end{document}